\documentclass[letterpaper,twocolumn,pra,
aps,showpacs,superscriptaddress,floatfix]{revtex4-2}

\usepackage[latin1]{inputenc}
\usepackage{bm}
\usepackage[usenames]{color}
\usepackage{multirow}
\usepackage{amssymb}
\usepackage{amsbsy}
\usepackage{mathtools}
\usepackage{amsmath}
\usepackage{graphicx}
\usepackage{epsfig}
\usepackage{placeins}
\usepackage{braket}
\usepackage{blindtext}
\usepackage[colorlinks,linkcolor=blue,citecolor=blue,urlcolor=blue]{hyperref}

\usepackage{scalerel}
\usepackage{tikz}
\newcommand{\C}[1]{{\cal{#1}}}
\newcommand{\bs}[1]{\boldsymbol{#1}}
\newcommand{\new}[1]{{#1}}

\usetikzlibrary{calc}
\usetikzlibrary{patterns}
\usetikzlibrary{svg.path}
\definecolor{orcidlogocol}{HTML}{A6CE39}
\tikzset{
	orcidlogo/.pic={
		\fill[orcidlogocol] svg{M256,128c0,70.7-57.3,128-128,128C57.3,256,0,198.7,0,128C0,57.3,57.3,0,128,0C198.7,0,256,57.3,256,128z};
		\fill[white] svg{M86.3,186.2H70.9V79.1h15.4v48.4V186.2z}
		svg{M108.9,79.1h41.6c39.6,0,57,28.3,57,53.6c0,27.5-21.5,53.6-56.8,53.6h-41.8V79.1z M124.3,172.4h24.5c34.9,0,42.9-26.5,42.9-39.7c0-21.5-13.7-39.7-43.7-39.7h-23.7V172.4z}
		svg{M88.7,56.8c0,5.5-4.5,10.1-10.1,10.1c-5.6,0-10.1-4.6-10.1-10.1c0-5.6,4.5-10.1,10.1-10.1C84.2,46.7,88.7,51.3,88.7,56.8z};
	}
}

\newcommand\orcid[1]{\!%
	\href{https://orcid.org/#1}{%
		\mbox{%
			\scaleto{%
				\begin{tikzpicture}[yscale=-1,transform shape]
				\pic{orcidlogo};
				\end{tikzpicture}
			}{8pt}%
		}%
	}%
}

\makeatletter

\newcommand{\be}{\begin{equation}}
\newcommand{\ee}{\end{equation}}

\makeatletter

\begin{document}
	\title{Decoherence of Histories: Chaotic Versus Integrable Systems}
	\date{\today}
	\author{Jiaozi Wang~\orcid{0000-0001-6308-1950}}
		\email{jiaowang@uos.de}
	\affiliation{Department of Mathematics/Computer Science/Physics, University of Osnabr\"uck, D-49076 
		Osnabr\"uck, Germany}

	\author{Philipp Strasberg~\orcid{0000-0001-5053-2214}}
	\affiliation{F\'isica Te\`orica: Informaci\'o i Fen\`omens Qu\`antics, Departament de F\'isica, Universitat Aut\`onoma de Barcelona, 08193 Bellaterra (Barcelona), Spain}
	\affiliation{Instituto de F\'isica de Cantabria (IFCA), Universidad de Cantabria--CSIC, 39005 Santander, Spain}

	\begin{abstract}
		We study the emergence of decoherent histories in isolated systems based on exact numerical integration of the Schr\"odinger equation for a Heisenberg chain. We reveal that the nature of the system, which we switch from (i) chaotic to (ii) interacting integrable to (iii) non-interacting integrable, strongly impacts decoherence \new{of coarse spin observables}. From a finite size scaling law we infer a strong exponential suppression of coherences for (i), a weak exponential suppression for (ii) and no exponential suppression for (iii) on a relevant short (nonequilibrium) time scale. Moreover, for longer times we find stronger decoherence for (i) but the opposite for (ii), hinting even at a possible power-law decay for (ii) at equilibrium time scales. This behaviour is encoded in the multi-time properties of the quantum histories and it can not be explained by environmentally induced decoherence. Our results suggest that chaoticity plays a crucial role in the emergence of classicality in finite size systems.
	\end{abstract}
	
	\maketitle
	
	{\it Introduction.---}\new{Decoherent histories} play a crucial role to explain the emergence of classicality \new{in isolated quantum systems}~\cite{GellMannHartleInBook1990, OmnesRMP1992, HalliwellANY1995, DowkerKentJSP1996}. \new{Mathematically, the diagonality of} the decoherence functional (DF)---quantifying the (de)coherence between different paths or histories in an isolated quantum system (precisely defined below)--- provides the condition for the existence of records of \emph{past} events~\cite{AlbrechtPRD1992, GellMannHartlePRD1993, HalliwellPRD1999, DoddHalliwellPRD2003, RiedelZurekZwolakPRA2016, HartleArXiv2016}, which may be used to identify branches in the universal wave function~\cite{GellMannHartleInBook1990, StrasbergReinhardSchindlerPRX2024}. It also implies \new{the validity of} Leggett-Garg inequalities~\cite{LeggettGargPRL1985, EmaryLambertNoriRPP2014} \new{and allows to replace} a complex quantum process by a simpler classical stochastic process~\cite{GriffithsJSP1984, SmirneEtAlQST2018, StrasbergDiazPRA2019, MilzEtAlQuantum2020, MilzEtAlPRX2020, StrasbergEtAlPRA2023, StrasbergSP2023, SzankowskiCywinskiQuantum2024}.
	
	Historically, since the DF is a complicated multi-time correlation function, research has been restricted to evaluating it explicitly for the case of quantum Brownian motion only~\cite{SchmidAP1987, DowkerHalliwellPRD1992, GellMannHartlePRD1993, HalliwellPRD1999, BrunHartlePRD1999, HalliwellPRD2001, SubasiHuPRE2012}. This non-interacting integrable model has been solved for a bath prepared in a canonical ensemble, a state that is highly mixed and contains (in the conventionally considered continuum limit) an infinite amount of classical noise. This leaves it unclear whether the microscopic origin of the observed decoherence is an intrinsic feature of the dynamics or an artifact of the classical initial state. Also other indirect arguments for emergent decoherent histories were based on linear oscillator chains~\cite{HalliwellPRD2003}. In addition, the thermodynamic limit makes the question how exactly decoherence emerges as a function of system size inaccessible.
	
	The main contribution of this letter is an approximation-free evaluation of the DF for a realistic many-body system (a Heisenberg chain) based on exact numerical integration of the Schr\"odinger equation \cite{Numerical-Method} without using classical ensembles, thereby significantly extending a few previous studies evaluating the DF for pure states and finite size systems numerically~\cite{BrunHalliwellPRD1996, GemmerSteinigewegPRE2014, SchmidtkeGemmerPRE2016, NationPorrasPRE2020, AlbrechtBaunachArrasmithPRD2022, StrasbergEtAlPRA2023, StrasbergSP2023, StrasbergReinhardSchindlerPRX2024, StrasbergSchindlerArXiv2023} or on a quantum computer~\cite{arrasmith2019variational}. By leveraging the power of modern computers we extract a finite size scaling law and reveal that the nature of the system (chaotic, interacting integrable or free) strongly influences the emergence of decoherence at least for finite size systems. Our results support van Kampen's \new{old idea that coarse and slow observables of chaotic quantum systems are described by classical stochastic processes~\cite{VanKampenPhys1954}. Interestingly, while it has been argued that coarse and slow (or ``quasi-conserved'') observables are important for decoherent histories~\cite{GellMannHartlePRD1993, HalliwellPRD1998, HalliwellPRL1999, GellMannHartlePRA2007}, the influence of chaos has never been considered.}
	
	Our results also illuminate the debated relation to environmentally induced decoherence (EID)~\cite{JoosEtAlBook2003, ZurekRMP2003, SchlosshauerPR2019}, which is mathematically \emph{not} equivalent to decoherent histories (for detailed studies see Refs.~\cite{KieferCQG1991, AlbrechtPRD1992, FinkelsteinPRD1993, PazZurekPRD1993, RiedelZurekZwolakPRA2016, StrasbergSP2023}). In our model, the relevant reduced density matrix exactly commutes with the relevant observable at all times, yet the histories are not exactly decoherent as one might naively expect. Since the block-diagonal form of the reduced density matrix is here caused by symmetry, this does not contradict the idea of EID, but it illustrates how subtle the relation is: it is a clear-cut example for emergent decoherent histories that are not caused by the entanglement between subsystems. Moreover, while (single-particle) chaos was found to be an obstacle for the quantum-to-classical transition, which can be cured by EID~\cite{ZurekPazPRL1994, ZurekPS1998, SchlosshauerFP2008, CalzettaCQG2012, HernandezRanardRiedelArXiv2023}, this letter reveals that (many-body) chaos \emph{significantly} enhances the emergence of decoherent histories, highlighting an intriguing dual role of chaos.

	More broadly seen, our letter contributes to a deeper understanding of complex quantum dynamics beyond single time expectation values and reduced density matrices. While most current research debates the use of Loschmidt echos~\cite{GorinEtAlPR2006}, out-of-time-order correlators~\cite{SwingleNP2018} or process tensors~\cite{MilzModiPRXQ2021} as a diagnostic tool of quantum chaos (see, e.g., Refs.~\cite{CucchiettiEtAlPRL2003, KitaevKITP2015, MaldacenaShenkerStanfordJHEP2016, OTOC-Galitski17, MurthySrednickiPRL2019, XuScaffidiCaoPRL2020, SanchezEtAlPRA2022, DowlingKosModiPRL2023, DowlingModiPRXQ2024}), our results suggest quantum histories as another sensitive tool. Our example illustrates that histories contain crucial information about the nature of the system that is not revealed in the dynamical behaviour of single-time expectation values.
	
	{\it Preliminaries.---}We consider an isolated quantum system with Hamiltonian $H$, Hilbert space ${\cal H}$ with dimension $D=\dim\C H$ and initial state $|\psi_0\rangle$. We divide the Hilbert space $\C H = \bigoplus_x\C H_x$ into orthogonal subspaces $\C H_x$ corresponding to a complete set of orthogonal projectors $\{\Pi_x\}_{x=1}^M$ satisfying $\Pi_{x}\Pi_{y}=\delta_{x,y}\Pi_{x}$ and $\sum_{x=1}^{M}\Pi_{x}=I$ (with $I$ the identity). We call the set $\{\Pi_x\}$ a \emph{coarse-graining} because the emergence of decoherence requires $\Pi_x$ to belong to a coarse observable $A = \sum_{x=1}^{M}a_{x}\Pi_{x}$ characterized by large subspace dimensions $V_x \equiv \dim\C H_x \gg 1$.
	
	In spirit of a generalized Feynman path integral, we now write the unitary evolution of the wave function as a sum over histories
	\be\label{eq-psi-depo}
	|\psi_{n}\rangle=\sum_{x_{n}}\cdots\sum_{x_{1}}\sum_{x_{0}}|\psi(x_{n},\dots,x_{1},x_{0})\rangle,
	\ee
	where $(x_n,\dots,x_1,x_0)$ denotes a history corresponding to a state passing through subspaces $x_k$ at times $t_k$: 
	\be
	|\psi(x_{n},\cdots,x_{1},x_{0})\rangle\equiv\Pi_{x_{n}}U_{n,n-1}\cdots\Pi_{x_{1}}U_{1,0}\Pi_{x_{0}}|\psi_{0}\rangle.
	\ee
	Here, $U_{k,i} = e^{-iH(t_k - t_i)}$ is the unitary time evolution operator from time $t_i$ to $t_k$ ($\hbar\equiv1$). For brevity, we denote a history as $\bs x = (x_n,\dots,x_1,x_0)$ such that Eq.~\eqref{eq-psi-depo} becomes $|\psi_n\rangle=\sum_{\boldsymbol{x}}|\psi(\boldsymbol{x})\rangle$. Moreover, the length of a history is $L=n+1$. 
	
	The central object of study in the following is the decoherence functional (DF)~\cite{GellMannHartleInBook1990, OmnesRMP1992, HalliwellANY1995, DowkerKentJSP1996}
	\be
	{\mathfrak D}(\boldsymbol{x};\boldsymbol{y})\equiv\langle\psi(\boldsymbol{y})|\psi(\boldsymbol{x})\rangle,
	\ee
	which quantifies the overlap, or interference, between different histories $\bs x$ and $\bs y$. Owing to $\Pi_x \Pi_y = \delta_{x,y}\Pi_x$, it is true that $\mathfrak{D}(\boldsymbol{x};\boldsymbol{y})\sim\delta_{x_{n},y_{n}}$, i.e, the DF is always ``diagonal'' with respect to the final points of the history, but the DF is usually \emph{not} diagonal with respect to earlier times $t_{n-1},\cdots,t_0$ of the history. The special case where
	\be\label{eq-DHC}
	\mathfrak{D}(\boldsymbol{x};\boldsymbol{y})=0\text{ for all }\boldsymbol{x}\neq\boldsymbol{y}
	\ee
	is known as the {\it decoherent histories condition} (DHC). Then, only the diagonal elements of the DF survive, which equal the probability $\mathfrak{D}(\boldsymbol{x};\boldsymbol{x})$ to get measurement outcomes $\boldsymbol{x}$ according to Born's rule.
	
	In reality, for finite systems Eq.~\eqref{eq-DHC} only strictly holds in trivial cases, e.g., when the projectors commute with the time evolution operator. Usually, the off-diagonal elements of the DF are non-zero complex numbers and it becomes more appropriate to quantify the amount of (de)coherence between histories via~\cite{DowkerHalliwellPRD1992} 
	\be\label{eq-DFnormalized}
	\epsilon(\boldsymbol{x};\boldsymbol{y})\equiv\frac{\mathfrak{D}(\boldsymbol{x};\boldsymbol{y})}{\sqrt{\mathfrak{D}(\boldsymbol{x};\boldsymbol{x})\mathfrak{D}(\boldsymbol{y};\boldsymbol{y})}}.
	\ee
	We then have $|\epsilon(\bs x;\bs y)| \le 1$ (by Cauchy-Schwarz) such that an appropriate notion of decoherence arises for $|\epsilon(\bs x;\bs y)| \ll 1$. The central objective of this letter is to study the decay of $\epsilon(\bs x;\bs y)$ as a function of the particle number $N$ for different classes of systems discussed more precisely below: non-integrable (or chaotic), interacting integrable and non-interacting integrable (or free).
	
	Since it is cumbersome to study $\epsilon(\bs x;\bs y)$ for every pair of histories $(\bs x;\bs y)$, we consider two quantities. First, we quantify the average amount of decoherence by 
	\be
	\overline{\epsilon}=\frac{1}{M^{2L-1}-M^{L}}\sum_{\boldsymbol{x}\neq\boldsymbol{y}}|\epsilon(\boldsymbol{x};\boldsymbol{y})|,
	\ee
	where $M^{2L-1}-M^{L}$ equals the number of non-trivial pairs $(\bs x;\bs y)$ (excluding those for which $\bs x = \bs y$ and $x_n \neq y_n$). Second, statistical outliers and the worst case scenario (the maximum coherence between histories) are captured by
	\be
	\epsilon_{\text{max}}=\max_{\boldsymbol{x}\neq\boldsymbol{y}}|\epsilon(\boldsymbol{x};\boldsymbol{y})|.
	\ee

	\begin{figure}[t]
		\includegraphics[width=1\columnwidth]{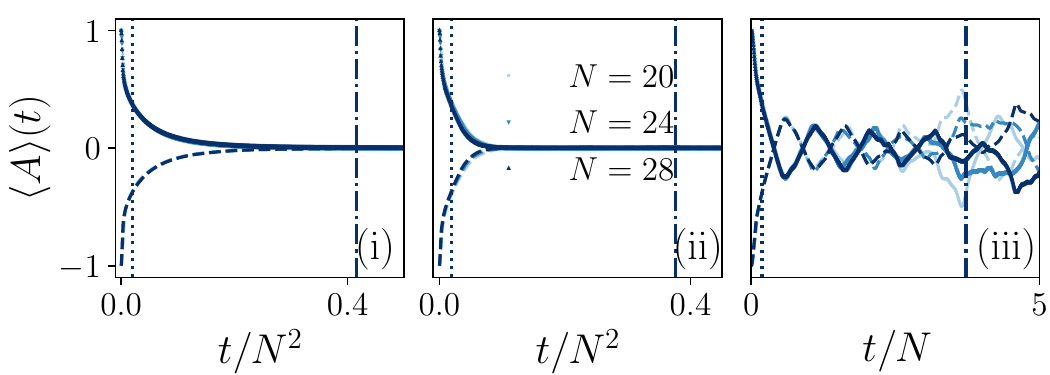}
		\caption{$\langle A\rangle(t)$ as a function of rescaled time for initial states $|\psi^+_0\rangle$ (solid line) and $|\psi^-_0\rangle$ (dashed line) for the chaotic (i), interacting integrable (ii) and non-interacting integrable (iii) cases \new{for increasing system size (light to dark blue)}. The dotted (dash-dotted) vertical line represents $T= T_\text{neq}$ ($T = T_\text{eq}$), where $T_\text{neq}$ indicates the time at which $\langle A\rangle(t)$ decays to $e^{-1}$ of its initial value and $T_\text{eq} = 20 T_\text{neq}$.
		}
		\label{Pt}
	\end{figure}

	{\it Model.---}As a paradigmatic quantum many-body system we consider a XXZ Heisenberg spin chain with Hamiltonian
	\be\label{eq-H}
	H=\sum_{\ell=1}^{N} \left(s_{x}^{\ell}s_{x}^{\ell+1}+s_{y}^{\ell}s_{y}^{\ell+1}+\Delta_{1}s_{z}^{\ell}s_{z}^{\ell+1}+\Delta_{2}s_{z}^{\ell}s_{z}^{\ell+2}\right),
	\ee
	where $s^\ell_{x,y,z} = \sigma^\ell_{x,y,z}/2$ are spin operators at lattice sites $\ell$, $N$ is the length of the chain \new{(chosen to be even)}, and we assume periodic boundary conditions. Crucial for our purposes is that Eq.~(\ref{eq-H}) contains three classes of systems for different parameter regimes: (i) for $\Delta_1\neq0\neq\Delta_2$ (we choose $\Delta_1 = 1.5$, $\Delta_2 = 0.5$) the model is non-integrable (or chaotic) meaning that nearest-level-spacing follows a Wigner-Dyson distribution \cite{RichterEtAlPRE2020}, and it satisfies the eigenstate thermalization hypothesis \cite{DAlessioEtAlAP2016, DeutschRPP2018};
	(ii) for $\Delta_1\neq 0$ but $\Delta_2 = 0$ (we choose $\Delta_1 = 1.5$) it is an interacting integrable model, which is solvable by Bethe ansatz~\cite{Bethe1931}; (iii) for $\Delta_1 = \Delta_2 = 0$ the model is non-interacting integrable (or free) and can be mapped to a quadratic Hamiltonian (a set of free fermions) via Jordan-Wigner transformation~\cite{book-xxz}. The fact that (ii) is integrable but can not be mapped to free fermions like (iii) is crucial: it qualitatively influences its transport behavior~\cite{Transport-RMP}, operator complexity~\cite{Parker19} and, as we reveal below, its decoherence.

	As an interesting observable we study the spin-imbalance operator,
	\be
	A_0= S_z^L - S_z^R = \sum_{\ell=1}^{\frac{N}{2}}s_{z}^{\ell}-\sum_{\ell=\frac{N}{2}+1}^{N}s_{z}^{\ell},
	\ee
	which quantifies a ``magnetization bias'' between the left and right half of the spin chain. Denoting its eigenvectors and eigenvalues by $A_0|a_k\rangle = a_k|a_k\rangle$, we construct a coarse observable $A = \Pi_{+}-\Pi_{-}$ with projectors
	\be
	\Pi_{+}=\sum_{a_{k}>0}|a_{k}\rangle\langle a_{k}|\quad\text{and}\quad\Pi_{-}=\sum_{a_{k}\le0}|a_{k}\rangle\langle a_{k}|.
	\ee
	As the total magnetization  $S_{z}=\sum_{\ell=1}^{N}s_{z}^{\ell}$ in $z$-direction commutes with $H$, we restrict the dynamics to a subspace with fixed $S_z$. We choose $S_z = 0$ for system size $N = 4k + 2$ with resulting Hilbert space dimension $D = \binom{N}{2k+1}$ and $S_z = 1$ for $N = 4k$ with $D = \binom{N}{2k+2}$, where $k\in \mathbb{N}$. In this way, we ensure equal subspace dimensions $V_+ = V_-$ with  $V_\pm = \dim\C H_\pm$.

	\begin{figure}[t]
		\includegraphics[width=1\columnwidth]{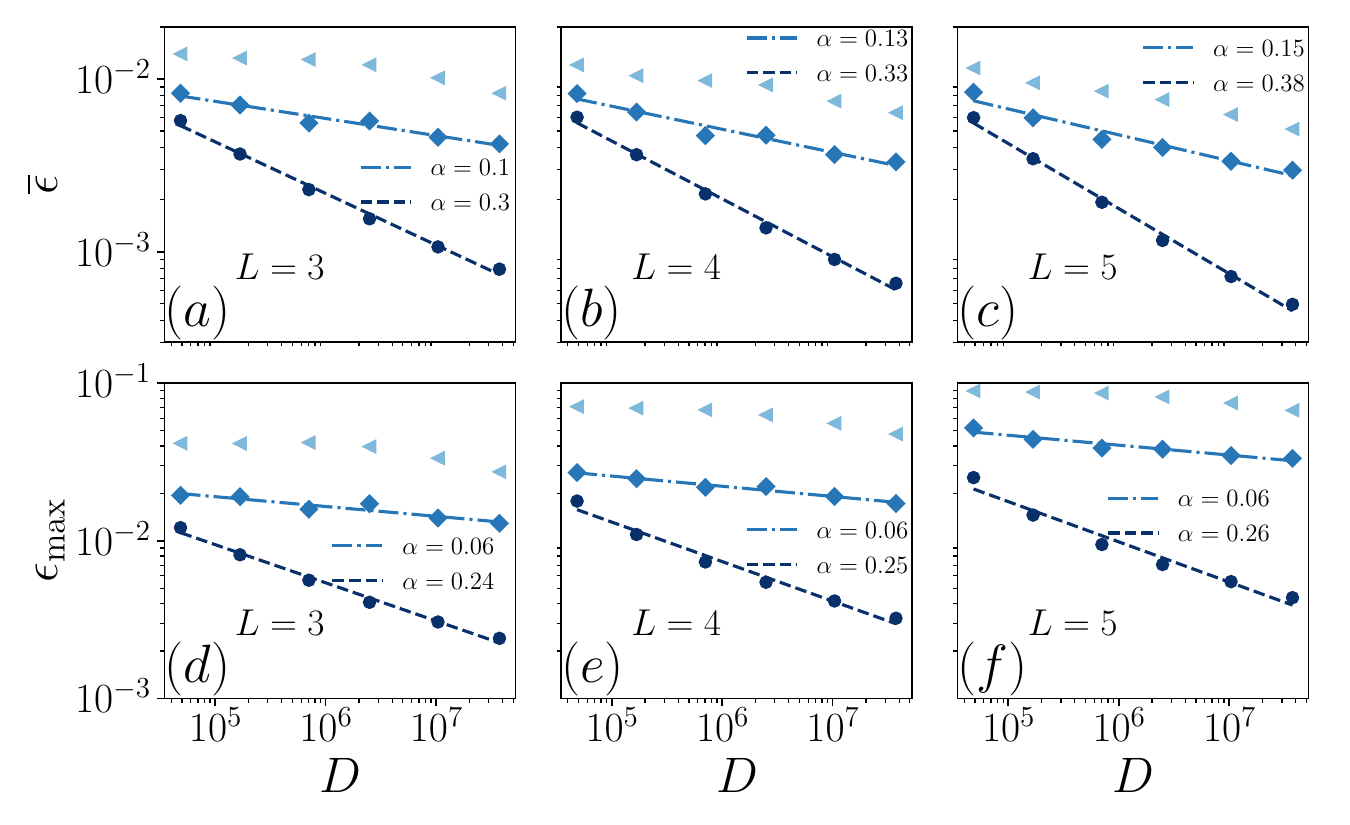}

		\caption{Average $\overline{\epsilon}$ and maximum $\epsilon_\text{max}$ amount of coherence versus Hilbert space dimension $D$ for the (i) chaotic (dark blue disks), (ii) interacting integrable (medium blue diamonds) and (iii) free (light blue triangles) case for $L \in\{3,4,5\}$. The dashed and dash-dotted line fit a scaling law of the form $D^{-\alpha}$ to (i) and (ii). The time step is $T = T_\text{neq}$ and the system sizes are $N=18,20,\dots, 28$.
		Note the double-logarithmic scale.
		}
		\label{DF-Short-Both}
	\end{figure}
	
	An interesting consequence of these choices is that the spin imbalance $A_0 = S_z^L - S_z^R$ can be determined by only measuring $S_z^L$ or $S_z^R$, owing to the conservation of $S_z = S_z^L + S_z^R$. Also owing to the conservation of $S_z$, the reduced density matrix of the left (right) half of the spin chain always commutes with $S_z^L$ ($S_z^R$), i.e., it is always block diagonal in the eigenbasis of $S_z^L$ ($S_z^R$). This is a consequence of symmetry and not of EID. Nevertheless, as we will see below, the histories are not exactly decoherent in that basis.

	{\it Numerical results.---}To get an overall picture of the average dynamics we plot the expectation value $\langle A\rangle(t)$ of the coarse spin imbalance as a function of time in Fig.~\ref{Pt} for two non-equilibrium initial states $|\psi_0^{\pm}\rangle$, where $|\psi_0^\pm\rangle$ is a Haar random state restricted to the subspace $\C H_\pm$. Remarkably, the behaviour in case (i) and (ii) \new{is very similar}: the system relaxes \new{approximately} exponentially to its thermal equilibrium value $\langle A\rangle_\text{eq} = 0$ with an equilibration time scale $\propto N^2$. In contrast, in case (iii) $\langle A\rangle(t)$ decays on a time scale $\propto N$ and fluctuates around $\langle A\rangle_\text{eq}$ without any clearly visible equilibration (up to the time that we considered).
	
\begin{figure}[t]
		\includegraphics[width=1\columnwidth]{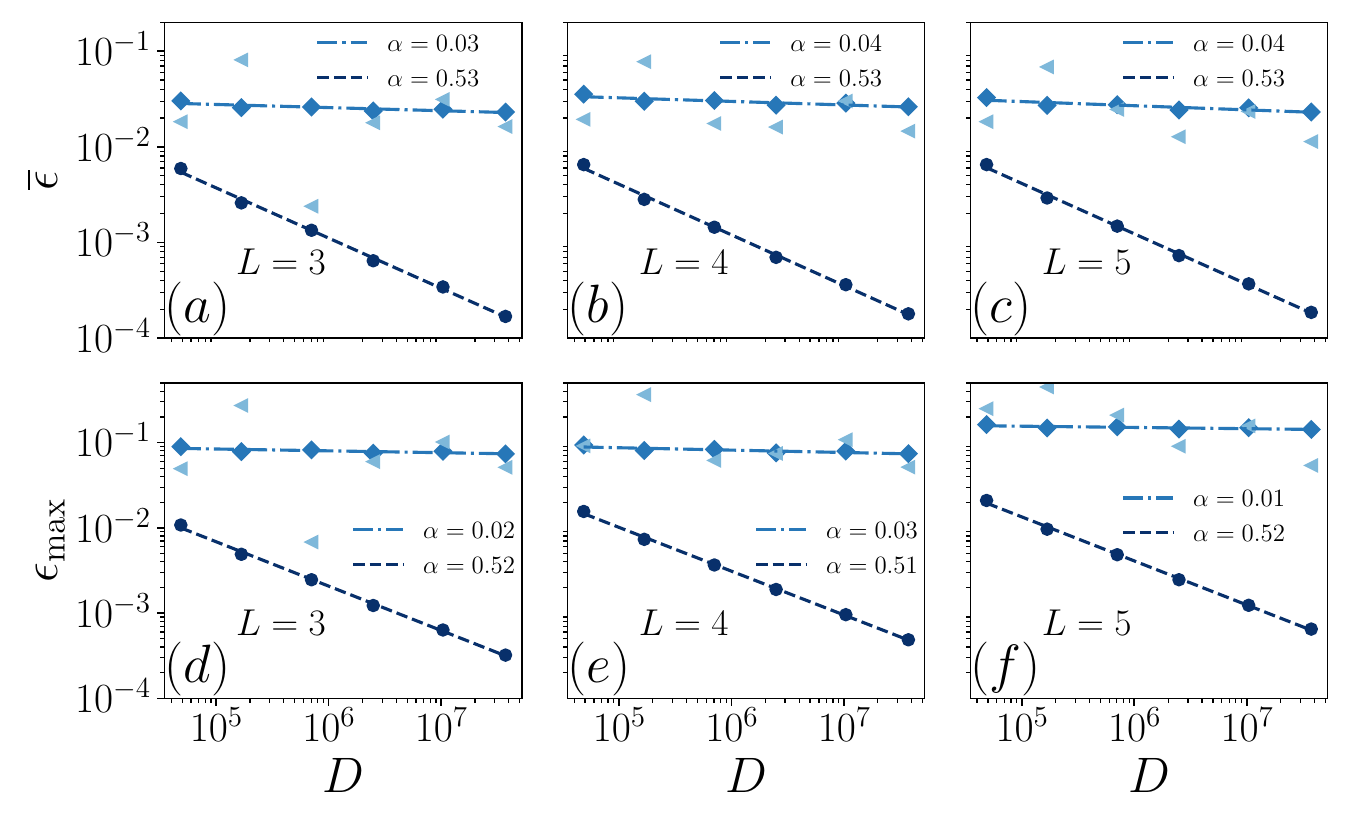}

		\caption{Identical to Fig. \ref{DF-Short-Both} except for time steps $T = T_\text{eq}$.
		}
		\label{DF-Long-Both}
	\end{figure}

	Decoherence is investigated in Figs.~\ref{DF-Short-Both} and \ref{DF-Long-Both} for Haar random initial states $|\psi_0\rangle$~\footnote{\new{We also considered initial product states and found results qualitatively similar to Figs.~\ref{DF-Short-Both} and~\ref{DF-Long-Both} (not shown here for brevity).}}. We plot in double logarithmic scale $\overline{\epsilon}$ and $\epsilon_\text{max}$ versus the Hilbert space dimension $D$ for histories of lengths $L\in\{3,4,5\}$; the case $L=2$ has a universal typical decay owing to the Haar random nature of the initial state as exemplified in the supplemental material (SM). The plots are obtained for constant time intervals $t_k - t_{k-1} = T$ for two different $T$: a nonequilibrium time scale $T_\text{neq}$ in Fig.~\ref{DF-Short-Both} (identical to the dotted line in Fig.~\ref{Pt}) and an equilibrium time scale $T_\text{eq}$ in Fig.~\ref{DF-Long-Both} (identical to the dash-dotted line in Fig.~\ref{Pt}). More precisely, $T_\text{neq}$ is defined as the time at which $\langle A\rangle (t)$ decays to $e^{-1}$ of its initial value \cite{Teq} and the equilibrium time scale is defined as $T_\text{eq} = 20 T_\text{neq}$. While in the free model (iii) there is no clearly visible equilibration in Fig.~\ref{Pt}, we use the same convention for $T_\text{eq}$ for better comparison. Each data point in Figs.~\ref{DF-Short-Both} and \ref{DF-Long-Both} is obtained by averaging $\overline{\epsilon}$ and $\epsilon_\text{max}$ over \new{$2^{32-N}$} different realizations of $|\psi_0\rangle$.  In the SM \new{we also study the fluctuations of $\overline{\epsilon}$ and $\epsilon_\text{max}$ as a function of $|\psi_0\rangle$, which do not influence our conclusions below. In particular, the standard deviation scales like $D^{-1/2}$, likely as a consequence of dynamical typicality~\cite{ReimannGemmerPA2020, TeufelTumulkaVogelJSP2023}.}
	
	We then observe in Figs.~\ref{DF-Short-Both} and \ref{DF-Long-Both} the following (an explanation follows later). First, for the chaotic case (i) we find a scaling law of the form $D^{-\alpha}$ (note that $D$ scales exponentially with the number of spins $N$) with $\alpha \approx 0.35$ ($\alpha\approx0.25$) for $\overline{\epsilon}$ ($\epsilon_\text{max}$) at the nonequilibrium time scale and with $\alpha\approx 0.5$ for both $\overline{\epsilon}$ and $\epsilon_\text{max}$ at the equilibrium time scale. This indicates a robust exponential suppression (with respect to system size $N$) of coherences in chaotic systems. For the interacting integrable case (ii) we find three major differences compared to (i). First, the exponent $\alpha$ is notably smaller in all cases. Second, for $T_\text{eq}$, $\alpha$ is smaller than for $T_\text{neq}$, conversely to case (i). Indeed, for $T_\text{eq}$ (Fig.~\ref{DF-Long-Both}) we even find $\alpha\approx0$, indicating a possible sub-exponential or power-law suppression (with respect to $N$) of coherences. Third, the exponent $\alpha$ for $\epsilon_\text{max}$ is roughly half the magnitude than $\alpha$ for $\overline{\epsilon}$ at $T_\text{neq}$, indicating stronger fluctuations in the DF among different pairs $(\bs x;\bs y)$ of histories. Finally, the free case (iii) is characterized for $T_\text{neq}$ by much larger coherences and for $T_\text{eq}$ by strong fluctuations among different $N$, making it unjustified to even speak of any scaling law $D^{-\alpha}$.

	\begin{figure}[t]
		\includegraphics[width=1\columnwidth]{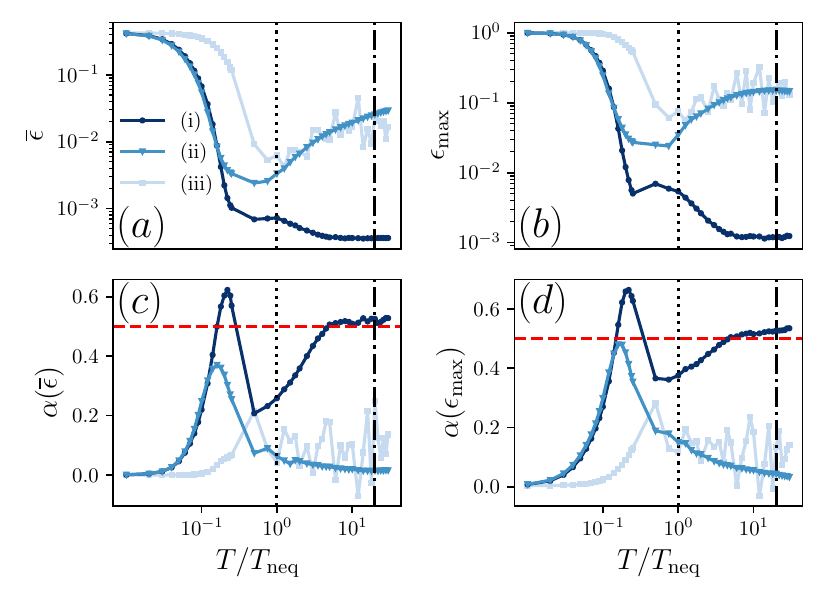}
		\caption{Decoherence as a function of the time step $T = t_k-t_{k-1}$ for history length $L=5$. (a,b) $\overline{\epsilon}, \epsilon_\text{max}$ for $N = 26$ on a double logarithmic scale. (c,d) Scaling exponent $\alpha$ for $\overline{\epsilon}, \epsilon_\text{max}$ obtained for system sizes $N = 18,20,\ldots,28$ and as a mean over \new{$2^{31-N}$} realizations of a Haar random $|\psi_0\rangle$.
		\new{Black dotted (dash-dotted) line indicates $T_\text{neq}$ $(T_\text{eq})$.}
			Red dashed line in (c,d) indicates $\alpha = 0.5$.
		}
		\label{DF-T-XXZ}
	\end{figure}
	
	A particularly intriguing observation is the distinctive behaviour of decoherence as a function of $T$, which is studied in detail in Fig.~\ref{DF-T-XXZ}. We first observe that decoherence is forbidden for $T\rightarrow0$ owing to the quantum Zeno effect. Still,  decoherence sets in on a short nonequilibrium time scale for (i) and (ii) and later for (iii). \new{What we found surprising is that for longer times $T\ge T_\text{neq}$ decoherence becomes weaker with increasing $T$ for (ii) and (iii) but not for (i). The scaling exponent $\alpha$ saturates to $\alpha \approx 0.5$ for (i), it decays to approximately zero for (ii), and it exhibits strong fluctuations for (iii).}
	
	{\it Explanation.---}We note that there is some consensus about the qualitative origin of decoherence (whether in the histories or EID framework), namely that coarse and slow observables of many-body systems decohere{, as considered here (in addition, we study in detail the connection to quasi-conserved hydrodynamic modes of spin density wave operators in the SM)}. However, other qualitative questions (e.g., is non-integrability essential or not?) have not been addressed in the past, and useful estimates of the DF are hard to obtain~\cite{StrasbergEtAlPRA2023, StrasbergSP2023}. Nevertheless, at least some quantitative features of the DF for chaotic systems seem to have a transparent explanation.
	
	To this end, recall that the overlap $\langle\phi|\chi\rangle$ between two Haar random vectors $|\phi\rangle$ and $|\chi\rangle$ scales like $1/\sqrt{D}$ and that the subspace dimensions $V_x$ are proportional to $D$ for many relevant coarse-grainings. Thus, for equilibrium time scales $T_\text{eq}$ the history states $|\psi(\bs x)\rangle$ in a chaotic system behave like randomly drawn typical states because they had time to explore the available Hilbert space in an unbiased way owing to the absence of conserved quantities (besides energy)\new{, and the irregularity of the time-dependent phases makes recoherences extremely unlikely for the vast majority of times}. Instead, for nonequilibrium time scales $T_\text{neq}$ the history states $|\psi(\bs x)\rangle$ will not look fully randomized as they contain further information compared to the equilibrium case. Therefore, one expects the exponent $\alpha = \alpha(T)$ to obey $\alpha(T)<0.5$ for $T<T_\text{eq}$, but Fig.~\ref{DF-T-XXZ} reveals exceptions for short time windows{, which we can not explain at the moment}. This indicates that the precise time dependence of $\alpha(T)$ is determined by $H$, $\{\Pi_x\}$ and $|\psi_0\rangle$ in a complicated way.
	
	Unfortunately, it is even more complicated to explain the behaviour for cases (ii) and (iii). Certainly, owing to the extensive number of conserved quantities, the states $|\psi(\bs x)\rangle$ can not explore the available Hilbert space in an unbiased fashion, which causes deviations from the behaviour of typical states. Yet, \new{that} the exponent $\alpha(T)$ becomes even smaller for larger $T$ \new{appears counterintuitive}. 
	\new{We believe the reason is that the dynamics, which is determined by terms of the form $\exp(-i\omega_{jk} T)$ with $\omega_{jk}$ a spectral gap, is less sensitive to the fine structure of $\{\omega_{jk}\}$ for short times $T$. The interesting point is that in the borderline case (ii) this sensitivity is revealed by the DF (a higher order correlation functions) but not by $\langle A\rangle(t)$ (cf.~Fig.~\ref{Pt}).}
	
	{\it Conclusion.---}We numerically extracted finite size scaling laws for the DF of a realistic quantum many-body system in an approximation-free way and we revealed decisive differences depending on the nature of the system. The chaotic case (i) showed a strong and robust emergence of decoherence in contrast to the interacting integrable case (ii) with a quantitatively much weaker form of decoherence. Note that this difference could not have been guessed from the quantitatively almost identical single-time behaviour shown in Fig.~\ref{Pt}. A qualitative even weaker form of decoherence was observed for free models (iii), making it even hard to speak of any definite signature of decoherence. Clearly, our finite size \new{and model dependent} calculations do not directly invalidate conclusions obtained from free models in the thermodynamic limit~\cite{SchmidAP1987, DowkerHalliwellPRD1992, GellMannHartlePRD1993, HalliwellPRD1999, HalliwellPRD2001, SubasiHuPRE2012, HalliwellPRD2003}. \new{However,} for another model (energy exchanges in an Ising chain, studied in the SM) we also found \new{strong differences between the chaotic and free case, suggesting that great care is needed when assessing the decoherence of free models.}
	
	Our findings motivate the conjecture that the normalized DF in Eq.~\eqref{eq-DFnormalized} can be written in the chaotic case for slow and coarse observables as
	\begin{equation}\label{eq-conjecture}
	\epsilon(\bs{x};\bs{y}) = \delta_{\bs x,\bs y} + (1-\delta_{\bs x,\bs y}) \frac{r_{\bs x,\bs y}}{D^\alpha}. 
	\end{equation}
	Here, $D^{-\alpha}$ describes the overall scaling with an exponent $\alpha$ that depends on many details (initial state, considered time interval, etc.) but is often not much smaller than $0.5$. Moreover, the coefficients $r_{\bs x,\bs y}$ are of order one and do not depend on $D$. While they are in principle determined by $H$, $\{\Pi_x\}$ and $|\psi_0\rangle$, they depend on so many experimentally uncontrollable microscopic parameters such that they appear erratic and unpredictable (similar to the off-diagonal elements in the eigenstate thermalization hypothesis~\cite{DAlessioEtAlAP2016, DeutschRPP2018}), a point that we further support in the SM. The conjecture \eqref{eq-conjecture} is in unison with previous analytical estimates~\cite{StrasbergEtAlPRA2023, StrasbergSP2023} and scaling laws~\cite{StrasbergReinhardSchindlerPRX2024}, and it breaks down when the number of histories $M^L$ becomes of the order of the Hilbert space dimension $D$~\cite{StrasbergSchindlerArXiv2023}.
		
	Finally, the notable difference in the value $\alpha$ at equilibrium time scales (and its time dependence) between chaotic and integrable systems suggests that $\alpha$ can be used as an indicator of quantum chaos. A main advantage of it is the  applicability to system sizes beyond the reach of exact diagonalization, e.g., through real-time propagation methods such as the Chebyshev polynomial algorithm used here.

	{\it Outlook.---}Our results motivate further research in various directions. On a quantitative level, for instance, it remains open to understand the precise behaviour of $\alpha(T)$ (and in particular the peculiarities of integrable models) as well as the effect of long histories with $L\gg 1$, which was numerically inaccessible to us. On a qualitative level, it would be intriguing to find out whether other classes of systems (e.g., disordered or localized systems) and phases of matter (e.g., close to criticality or topological phases) also have such a strong influence on the behaviour of decoherence as seen here.
	
	{\it Acknowledgement.---}JW is financially supported by the Deutsche
	Forschungsgemeinschaft (DFG), under Grant No. 531128043, as well as under Grant
	No.\ 397107022, No.\ 397067869, and No.\ 397082825 within the DFG Research
	Unit FOR 2692, under Grant No.\ 355031190. PS is financially supported by ``la Caixa'' Foundation (ID 100010434, fellowship code LCF/BQ/PR21/11840014), the Ram\'on y Cajal program RYC2022-035908-I, the MICINN with funding from European Union NextGenerationEU (PRTR-C17.I1), by the Generalitat de Catalunya (project 2017-SGR-1127), the European Commission QuantERA grant ExTRaQT (Spanish MICIN project PCI2022-132965), and the Spanish MINECO (project PID2019-107609GB-I00) with the support of FEDER funds. Additionally, we greatly acknowledge computing time on the HPC3 at the University of Osnabr\"{u}ck, granted by the DFG, under Grant No. 456666331.

	\bibliography{references}

\begin{thebibliography}{75}
\bibitem [{Num()}]{Numerical-Method}%
  \BibitemOpen
  \href@noop {} {}\bibinfo {note} {In this paper we numerically integrate the
  Schr{\"o}dinger equation using the Chebyshev polynomial algorithm
  \cite{Raedt07-Chebysev, Jin10-Chebyshev} until convergence is
  reached.}\BibitemShut {Stop}%
\bibitem [{Teq()}]{Teq}%
  \BibitemOpen
  \href@noop {} {}\bibinfo {note} {Specifically, we obtain $T_\text{neq}$ from
  the data of the largest system size $(N_\text{max} = 28)$. Then, for smaller
  sizes $N$ we set
  $T_{\text{neq}}(N)=(N/N_{\text{max}})^{p}T_{\text{neq}}(N_{\text{max}})$ with
  $p=2$ for (i) and (ii) and $p = 1$ for (iii).}\BibitemShut {Stop}%
\end{thebibliography}
	
	
	
\clearpage
\newpage

\setcounter{figure}{0}
\setcounter{equation}{0}
\renewcommand{\thefigure}{S\arabic{figure}}
\renewcommand{\theequation}{S\arabic{equation}}

\section*{Supplemental Material}
	We here report on further numerical results that strengthen our main message. We start by supplementing additional results for the XXZ Heisenberg chain studied in the main text (subsection S1), continued by comparing the chaotic and free case for energy \new{imbalance} in an Ising chain (subsection S2), and finally we look at the level statistics to demonstrate that our models are chaotic (subsection S3).

\subsection*{S1. Additional numerical results in XXZ model}

%
%
%
%
\subsubsection*{S1.A. Results on spin imbalance operator}
\begin{figure}[b]
	\includegraphics[width=1\columnwidth]{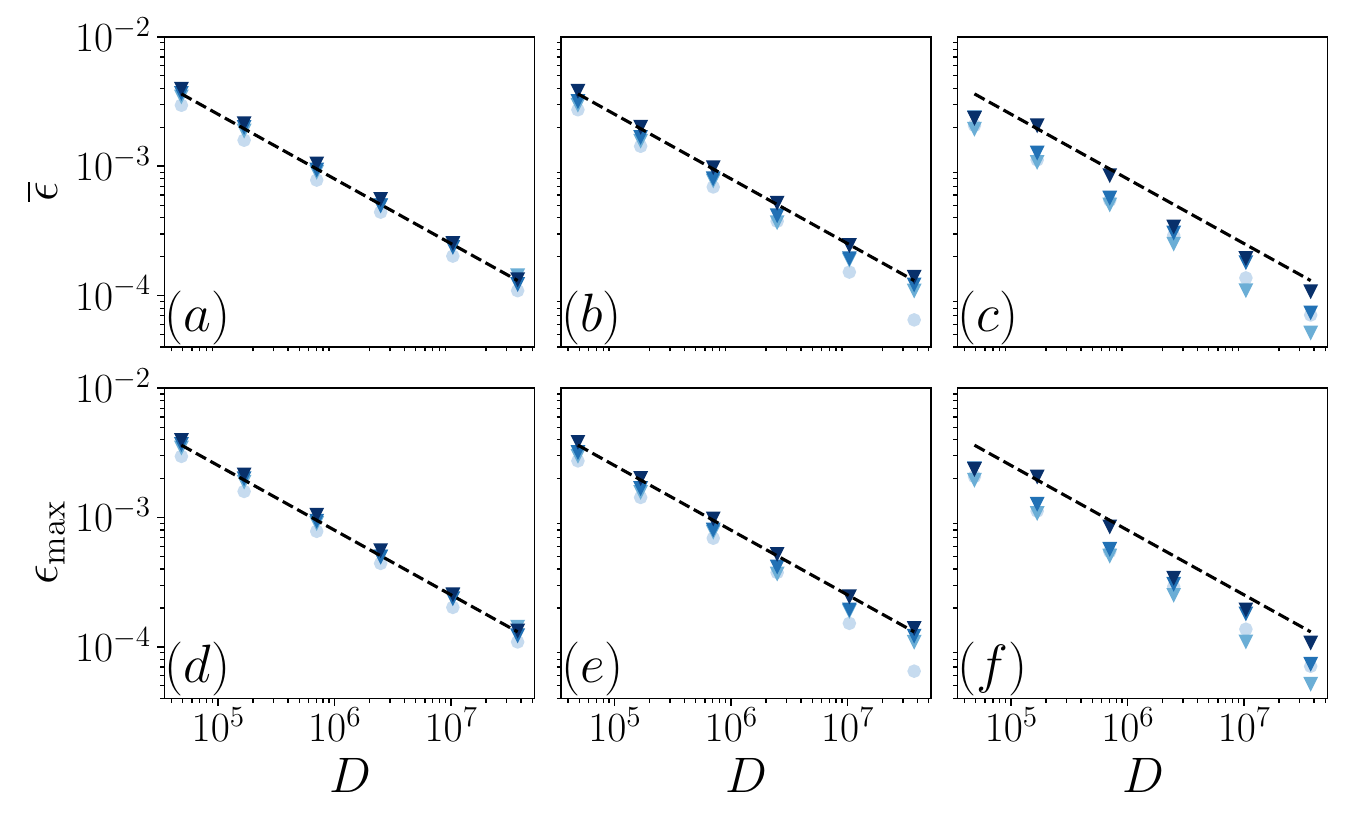}
	
	\caption{$\overline{\epsilon}$ and $\epsilon_\text{max}$ versus Hilbert space dimension $D$ for $L=2$, for chaotic (a,d), interacting integrable (b,e) and non-interacting integrable (c,f) cases for time steps $T=T_\text{neq},2T_\text{neq},3T_\text{neq},T_\text{eq}$ (from light to dark). The dashed line indicates the scaling $D^{-0.5}$ as a guidance to the eyes.
	}
	\label{DF-L2}
\end{figure}

We begin in Fig.~\ref{DF-L2} by displaying $\overline{\epsilon}$ and $\epsilon_\text{max}$ for the shortest possible non-trivial history with length $L=2$. As claimed in the main mansuscript, we observe consistently a $D^{-0.5}$ scaling in all three cases (i), (ii) and (iii). This effect is due to the Haar random choice of the initial state and the fact that the Haar measure is invariant under unitary transformations. Thus, the first unitary time evolution $U_{1,0}$ is barely able to change the decoherence properties of the system.

\begin{figure}[t]
	\includegraphics[width=1\columnwidth]{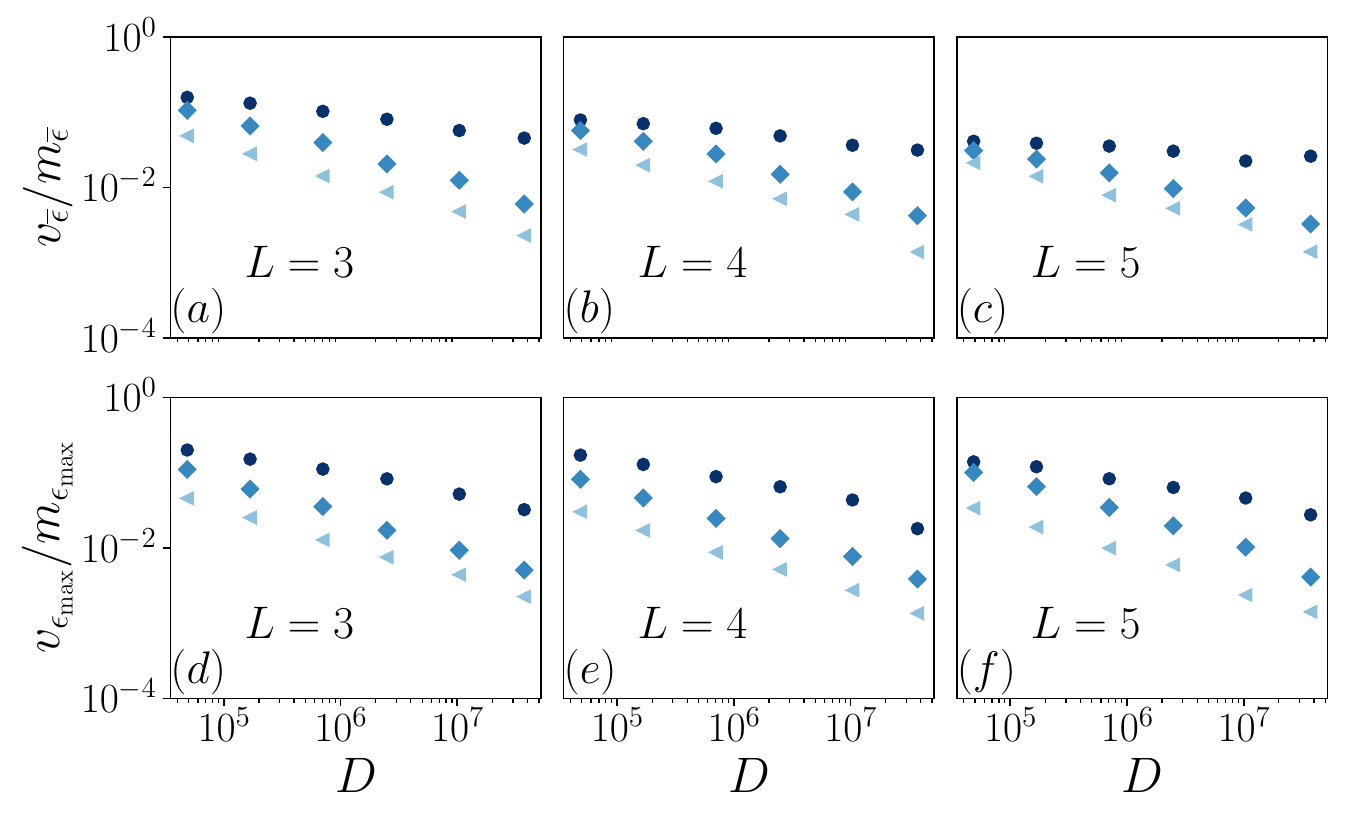}
	
	\caption{Inverse SNR versus Hilbert space dimension $D$ for cases (i) (dark blue disks), (ii) (medium blue diamonds) and (iii) (light blue triangles) for $L = 3,4,5$.
		The time step is $T = T_\text{neq}$ and the system sizes are $N=18,20,\dots,28$.
	}
	\label{V-Short-Both}
\end{figure}

\begin{figure}[t]
	\includegraphics[width=1\columnwidth]{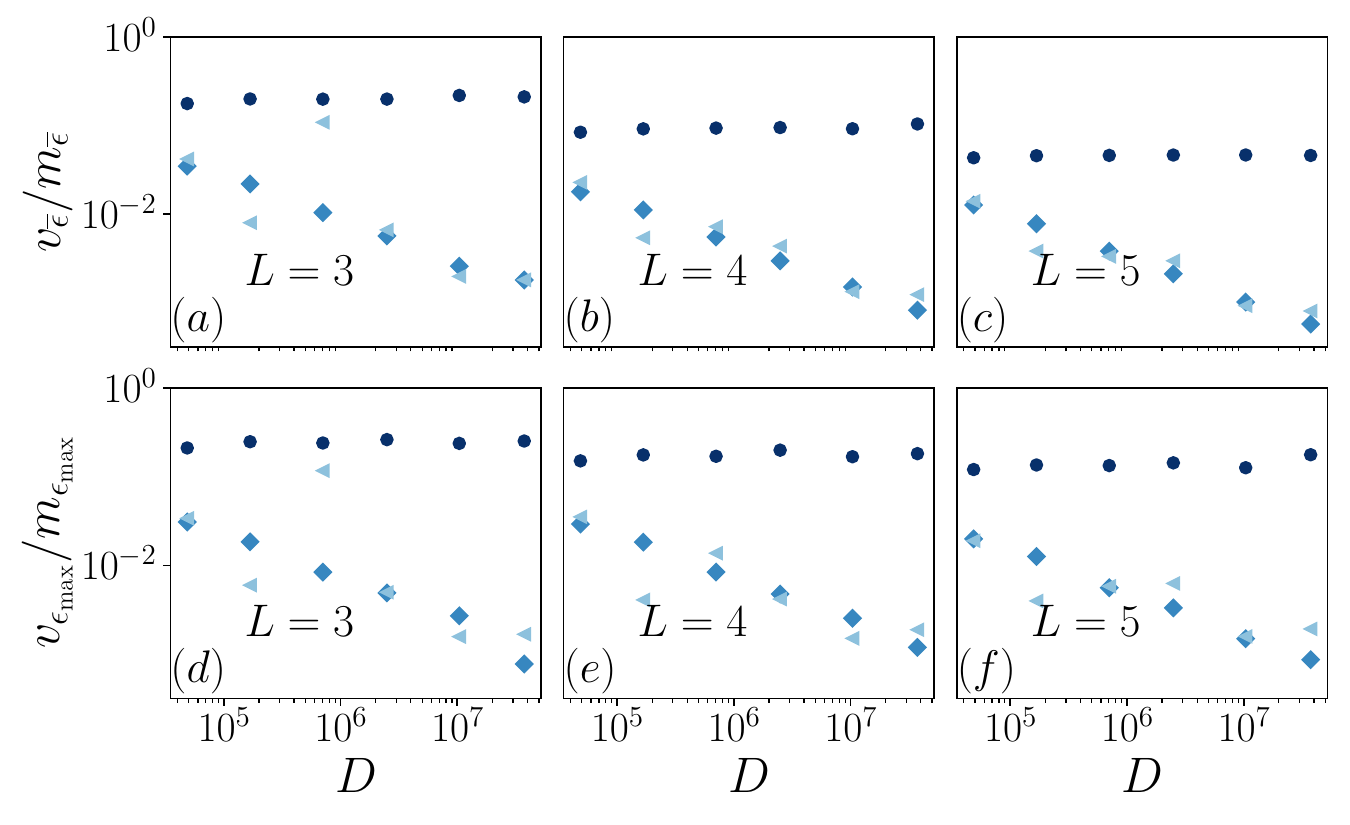}
	
	\caption{ Similar to Fig. \ref{V-Short-Both} but for the time scale $T = T_\text{eq}$.
	}
	\label{V-Long-Both}
\end{figure}

%

Next, we consider the question whether it is justified to focus on the averages of $\overline{\epsilon}$ and $\epsilon_\text{max}$ over the $2^{32-N}$ different realizations of $|\psi_0\rangle$ (denoted here as $m_{\overline{\epsilon}}$ and $m_{\epsilon_\text{max}}$), as done in the main text. To this end, we computed the standard deviation $v_{\overline{\epsilon}}$ and $v_{\epsilon_\text{max}}$, which we found to scale like $D^{-1/2}$ in all cases (i), (ii) and (iii) and for all $L\in\{3,4,5\}$ (likely as a consequence of dynamical typicality). This is not shown here for brevity because it is more revealing to consider the inverse signal-to-noise ratio (SNR) $v_{\overline{\epsilon}}/m_{\overline{\epsilon}} \sim D^{\alpha(\overline{\epsilon})-1/2}$ and $v_{\epsilon_\text{max}}/m_{\epsilon_\text{max}} \sim D^{\alpha(\epsilon_\text{max})-1/2}$ for $T_\text{neq}$ in Fig.~\ref{V-Short-Both} and for $T_\text{eq}$ in Fig.~\ref{V-Long-Both}. Since the integrable dynamics (ii) and (iii) is characterized by very weak decoherence (if at all), their inverse SNR quickly becomes vanishingly small. This is not so for the chaotic dynamics (i), where at equilibrium time scales $\alpha\approx 1/2$ implies no vanishing of the inverse SNR with system size. However, we can clearly see in Fig.~\ref{V-Long-Both} that the standard deviation is about an order of magnitude smaller than the mean value, justifying our focus on the average in the main text. Moreover, we also find that $v_{\overline{\epsilon}}/m_{\overline{\epsilon}}$ shows a clear decrease with $L$ because the size of the DF increases with $L$.

\begin{figure}[b]
	\includegraphics[width=0.9\columnwidth]{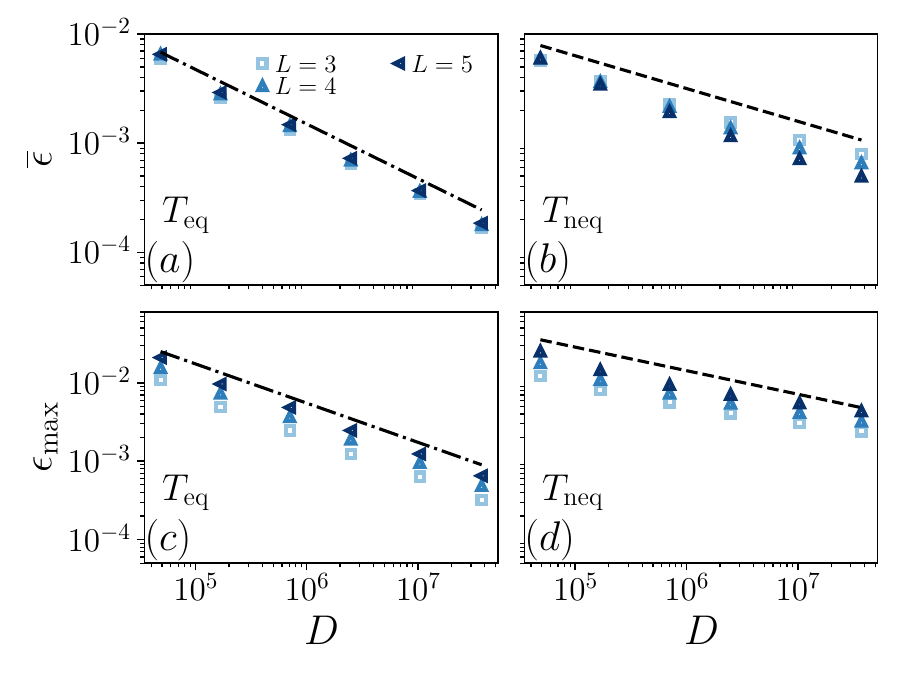}
	
	\caption{$\overline{\epsilon}$ and $\epsilon_\text{max}$ versus Hilbert space dimension $D$ for the chaotic case (i) for $L \in \{3,4,5\}$. The time step is $T = {T_\text{eq}}$ in (a,c) and $T = T_\text{neq}$ in (b,d). The dashed-dotted (dashed) line indicates the $\sim D^{-0.5}$ ($\sim D^{-0.3}$) scaling as a guidance to the eyes.
	}
	\label{DF-L}
\end{figure}

\new{Furthermore, we directly compare in Fig.~\ref{DF-L} how the average $\overline{\epsilon}$ and maximum coherence $\epsilon_\text{max}$ changes as a function of $L$ for the chaotic case. We observe in all four cases rather mild changes as a function of $L$, and there seems to be a convergence to a common value in unison with the scaling law extracted in the main text (however, for $L\gg1$, which was numerically not accessible for us, ``recoherences'' must eventually happen). Moreover, for $\epsilon_\text{max}$ we consistently observe a slight increase for larger $L$ for both $T_\text{eq}$ and $T_\text{neq}$. Since $\epsilon_\text{max}$ was defined to measure statistical outliers, this is likely caused by the fact that the DF has more entries for larger $L$ (the number of non-trivial entries grows like $M^{2L-1}-M^{L}$).}

\begin{figure}[t]
	\includegraphics[width=1.0\columnwidth]{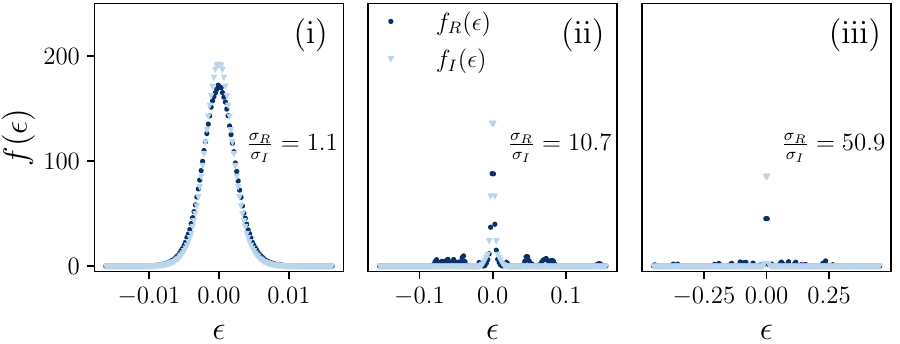}
	\caption{Distribution of real ($f_R(\epsilon)$) and imaginary ($f_I(\epsilon)$) part of ${\epsilon}_{\boldsymbol{x},\boldsymbol{y}}$ in the XXZ model for $N = 20$, $L = 5$ and $T = T_\text{eq}$ for (i) chaotic; (ii) interacting integrable and (iii) free cases. $\sigma_{R,I}$ indicates the standard deviation of $f_{R,I}(\epsilon)$. \new{Data from \new{$2^{12}$} different initial Haar random states are taken into account.}
	}
	\label{Fig-Dis}
\end{figure}

In addition to the averaged and maximum value of $\epsilon_{\boldsymbol{x},\boldsymbol{y}}$, we also study its distribution. As an example, in Fig.~\ref{Fig-Dis}, we show the distribution of the real and imaginary part of $\epsilon_{\boldsymbol{x},\boldsymbol{y}}$ for $N=20,\ L=5, \ T=T_\text{eq}$, where we exclude pairs of $\boldsymbol{x},\boldsymbol{y}$ for which $x_n \neq y_n$ or $x_0 \neq y_0$.
Data from \new{$2^{12}$} different initial Haar random states are taken into account. In the chaotic case (i), a similar shape of distribution is found for the real ($f_R(\epsilon)$) and imaginary part ($f_I(\epsilon)$). However, slight deviations are clearly observed, especially in the variance (second moments), indicating that $\epsilon_{\boldsymbol{x},\boldsymbol{y}}$ can not be random numbers in a strict sense. This is due to the fact that in a real system, starting from the same initial state, different histories are nevertheless correlated. However, in comparison with the integrable cases (ii) and (iii), where we observe large deviations between $f_R(\epsilon)$ and $f_I(\epsilon)$, the correlations in the chaotic case are almost negligible.

\subsubsection*{S1.B. Results on spin density wave operator}

\new{In a translation invariant system as we considered here, we introduce the spin density wave operator with wave number $q \in\{-N/2,\dots,0,\dots,N/2-1\}$ (as in the main text we consider $N$ to be an even number)
	\begin{equation}
	W_{q}^{0}=\sum_{\ell=1}^{N}\exp(-i\frac{2\pi q}{N}\ell)s_{z}^{\ell}.
	\end{equation}
	Note that $W_{q=0}^0 = \sum_{\ell=1}^{N}s_{z}^{\ell}$ is the total spin, which is conserved, and $W_{N-q}^{0}=(W_{q}^{0})^{\dagger}$. 
	The spin imbalance operator studied in the main text can be expressed as
	\begin{equation}
	A_0=\sum_{q=-N/2}^{N/2-1}c_{q}W^0_{q},
	\end{equation}
	where
	\begin{equation}\label{eq-cq}
	c_{q}=\begin{cases}
	\frac{1}{N}{\displaystyle \left(i\cot\left(\frac{\pi q}{N}\right)-1\right)} & \text{odd\ }q\\
	0 & \text{even\ }q
	\end{cases}.
	\end{equation}
	Using $A_0=A_0^\dagger=\sum_{q=-N/2}^{N/2-1}c_{q}^{*}(W_{q}^{0})^{\dagger}$, the auto-correlation function of $A_0$ can be written as
	\begin{equation}\label{eq-AAt}
	\langle A_{0}(t)A_{0}\rangle=\sum_{q,q^{\prime}=-N/2}^{N/2-1}c_{q^{\prime}}^{*}c_{q}\langle(W_{q^{\prime}}^{0}(t))^{\dagger}W_{q}^{0}\rangle,
	\end{equation}
	where $\langle\bullet\rangle:=\frac{1}{D}\text{Tr}[\bullet]$.
	To further simply Eq.~\eqref{eq-AAt}, let us introduce the translation operator $\cal T$, 
	\begin{equation}
	{\cal T}|\sigma_{1}\sigma_{2}\ldots\sigma_{L}\rangle=|\sigma_{L}\sigma_{1}\ldots\sigma_{L-1}\rangle,
	\end{equation}
	where $|\sigma_{1}\sigma_{2}\ldots\sigma_{L}\rangle$ is the simultaneous basis of $s^\ell_z$, and 
	$\sigma_\ell =\uparrow \downarrow$, indicating the spin up and down (along z-direction).
	It follows that ${\cal T}^{-1}s_{z}^{\ell}{\cal T}=s_{z}^{\ell-1}$ and
	\begin{align}
	{\cal T}^{-1}W_{q}^{0}{\cal T}&= e^{-2\pi iq/N} W_{q}^{0}, \nonumber \\
	{\cal T}^{-1}(W_{q}^{0})^{\dagger}{\cal T}&= e^{2\pi iq/N}(W_{q}^{0})^{\dagger} . \label{eq-Wq1}
	\end{align}
	Since the Hamiltonian is translational invariant ${\cal T}^{-1}H{\cal T}=H$, we find
	\begin{equation}
	\begin{split}
	\langle W_{q^{\prime}}^{0}(t)^{\dagger}W_{q}^{0}\rangle
	&= \langle{(\cal T}^{-1}W_{q^{\prime}}^{0}{\cal T})(t)^\dagger{\cal T}^{-1}W_{q}^{0}{\cal T}\rangle \\
	&= e^{2\pi i(q^\prime-q)/N} \langle(W_{q^{\prime}}^{0}(t))^{\dagger}W_{q}^{0}\rangle. \label{eq-Wq2}
	\end{split}
	\end{equation}
	This equation can only be satisfied if $\langle W_{q^{\prime}}^{0}(t)^{\dagger}W_{q}^{0}\rangle = 0$ for $q\neq q^\prime$ or $q=q^\prime$. Thus, Eq.~\eqref{eq-AAt} becomes
	\begin{equation}
	\langle A_{0}(t)A_{0}\rangle=\sum_{q=-N/2}^{N/2-1}|c_{q}|^{2}\langle W_{q}^{0}(t)^{\dagger}W_{q}^{0}\rangle,
	\end{equation}
	Assuming diffusive spin transport, which has been confirmed numerically for our cases $\Delta_1 = 1.5, \Delta_2 = 0.5$ and $\Delta_1 = 1.5, \Delta_2 = 0$ (see Ref. \cite{Transport-RMP} and references therein), we have
	\begin{equation}
	\langle W_{q}^{0}(t)^{\dagger}W_{q}^{0}\rangle\sim e^{-(\frac{2\pi q}{N})^{2}D_{c}t},\ \text{for}\ \frac{|q|}{N}\rightarrow0,
	\end{equation}
	where $D_c$ denotes the diffusion constant. For $N\rightarrow \infty$, Eq.~\eqref{eq-cq} becomes
	\begin{equation}
	|c_{q}|\simeq\begin{cases}
	{\displaystyle \frac{1}{\pi q}} & \text{odd}\ q,\ \frac{|q|}{N}\rightarrow 0\\
	0 & \text{otherwise}
	\end{cases}.
	\end{equation}
	As a result, the auto-correlation function of the spin imbalance operator $\langle A_0(t)A_0\rangle$ is dominated by the auto-correlation function of the density wave operator $\langle W^0_{q}(t)^\dagger W^0_{q}\rangle$ at small $q$, especially the late time behavior.}

\begin{figure}[t]
	\includegraphics[width=1\columnwidth]{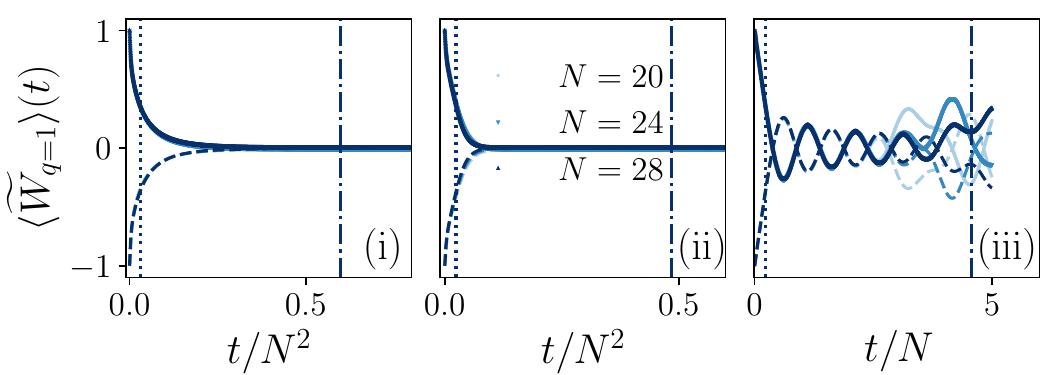}
	\caption{\new{$\langle \widetilde{W}_{q=1}\rangle(t)$ as a function of rescaled time for initial states $|\psi^+_0\rangle$ (solid line) and $|\psi^-_0\rangle$ (dashed line) for the chaotic (i), interacting integrable (ii) and non-interacting integrable (iii) cases. The dotted (dash-dotted) vertical line represents $T= T_\text{neq}$ ($T = T_\text{eq}$), where $T_\text{neq}$ indicates the time at which $\langle \widetilde{W}_{q=1}\rangle(t)$ decays to $e^{-1}$ of its initial value and $T_\text{eq} = 20 T_\text{neq}$. Specifically, we obtain $T_\text{neq}$ from the data of the largest system size $(N_\text{max} = 28)$. Then, for smaller sizes $N$ we set $T_{\text{neq}}(N)=(N/N_{\text{max}})^{p}T_{\text{neq}}(N_{\text{max}})$ with $p=2$ for (i) and (ii) and $p = 1$ for (iii).}
	}
	\label{Pt-DW}
\end{figure}

\begin{figure}[t]
	\includegraphics[width=1\columnwidth]{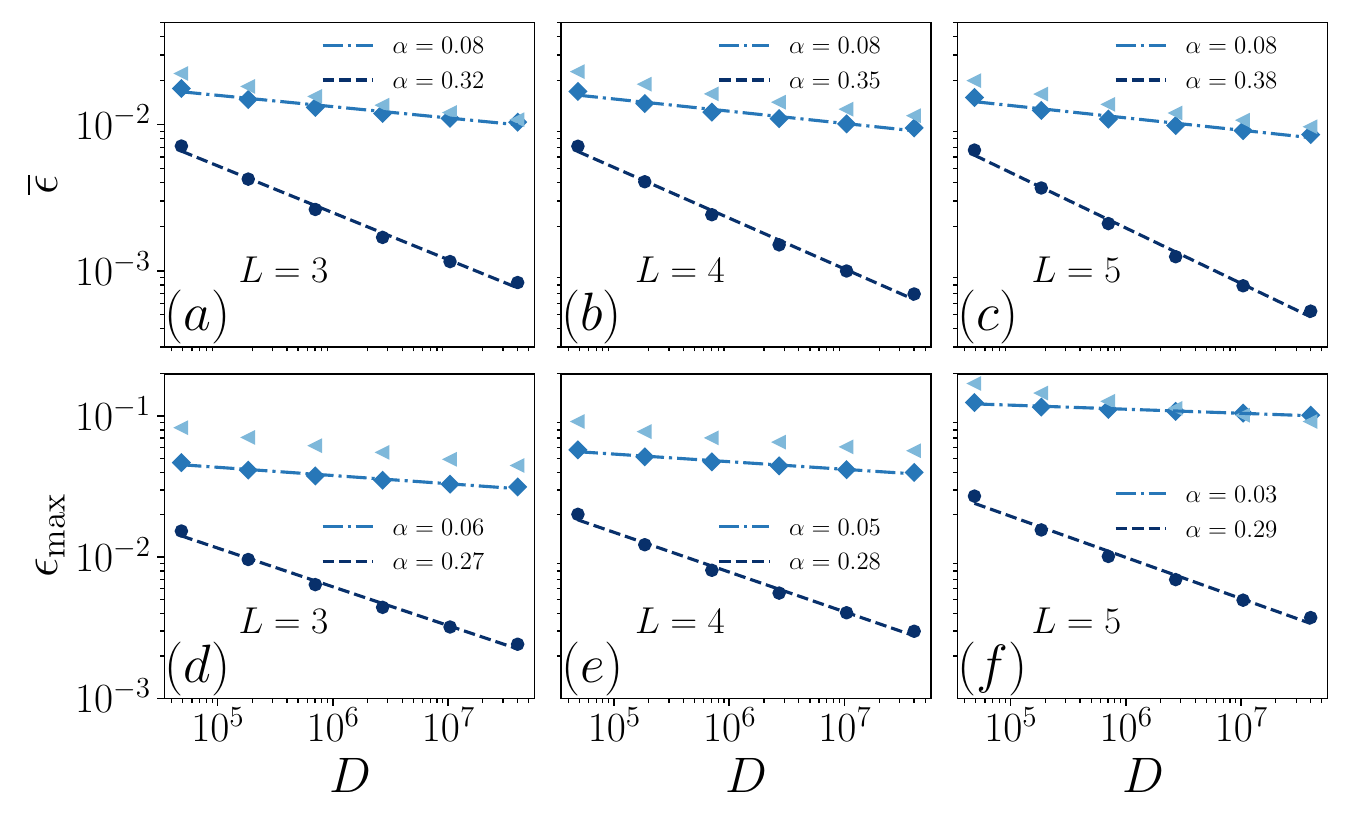}
	
	\caption{Average $\overline{\epsilon}$ and maximum $\epsilon_\text{max}$ amount of coherence versus Hilbert space dimension $D$ for the (i) chaotic (dark blue disks), (ii) interacting integrable (medium blue diamonds) and (iii) free (light blue triangles) case for $L \in\{3,4,5\}$ of the coarse spin density wave  $\widetilde{W}_{q=1}$. The dashed and dash-dotted line fit a scaling law of the form $D^{-\alpha}$ to (i) and (ii). The time step is $T = T_\text{neq}$ and the system sizes are $N=18,20,\dots, 28$. Note the double-logarithmic scale.
	}
	\label{DF-Short-DWS}
\end{figure}

\begin{figure}[b]
	\includegraphics[width=1\columnwidth]{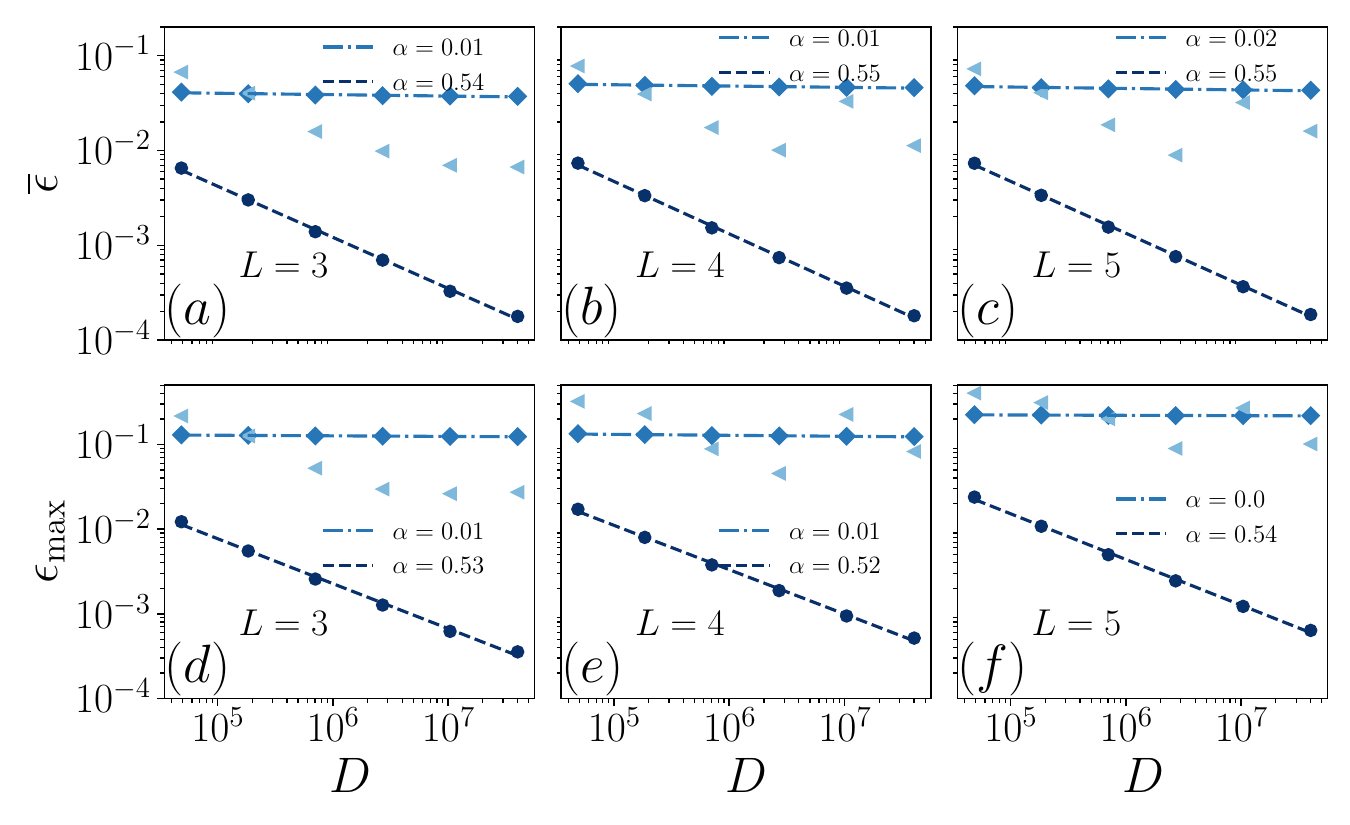}
	
	\caption{Identical to Fig. \ref{DF-Short-DWS} except for time steps $T = T_\text{eq}$.
	}
	\label{DF-Long-DWS}
\end{figure}

\begin{figure}[t]
	\includegraphics[width=1\columnwidth]{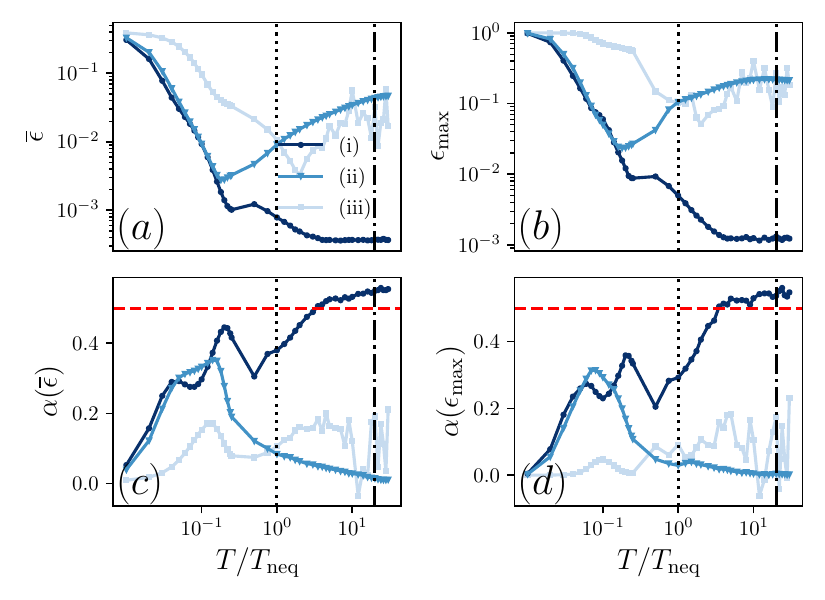}
	\caption{Decoherence as a function of the time step $T = t_k-t_{k-1}$ for history length $L=5$ of the coarse spin density wave $\widetilde{W}_{q=1}$. (a,b) $\overline{\epsilon}, \epsilon_\text{max}$ for $N = 26$ on a double logarithmic scale. (c,d) Scaling exponent $\alpha$ for $\overline{\epsilon}, \epsilon_\text{max}$ obtained for system sizes $N = 18,20,\ldots,28$ and as a mean over $2^{30-N}$ realizations of a Haar random $|\psi_0\rangle$. Red dashed line in (c,d) indicates $\alpha = 0.5$. Black dotted (dash-dotted) line indicates $T=T_\text{neq} (T=T_\text{eq})$.
	}
	\label{DF-T-XXZ-DWS}
\end{figure}

\new{In the following, we first consider the DF corresponding to the density wave operator $W^0_q$ for the slowest (non-trivial) mode $q=1$. For simplicity, we also consider the Hermitian version of it,
	\begin{equation}
	\widetilde{W}_{q}^{0}=\sum_{\ell=1}^{N-1}\cos(\frac{2\pi q}{N}\ell)s_{z}^{\ell}.
	\end{equation}
	The coarse grained operator can be constructed as, $\widetilde{W}_q = \Pi^W_{+}-\Pi^W_{-}$ with projectors
	\be
	\Pi^W_{+}=\sum_{w_{k}>0}|w_{k}\rangle\langle w_{k}|\quad\text{and}\quad\Pi^W_{-}=\sum_{w_{k}\le0}|w_{k}\rangle\langle w_{k}|,
	\ee
	where $w_k$ and $|w_k\rangle$ denote the eigenvalue and eigenvectors of $\widetilde{W}^0_{q=1}$, respectively. For all system sizes considered here, we focus on the subspace with $S_z = 0$.}

\new{Results for $\widetilde{W}_{q=1}$ are shown in Figs.~\ref{Pt-DW},~\ref{DF-Short-DWS},~\ref{DF-Long-DWS} and~\ref{DF-T-XXZ-DWS}, which are analogous to Figs.~1, 2, 3 and 4 of the main text. We consistently observe the same qualitative behavior as in the main text with small quantitative differences. In particular, we again observe strong differences between the  (i) chaotic, (ii) interacting integrable and (iii) free case. This clearly indicates that the results of the main text are general.}

\new{We remark that this nicely ties together the ideas of van Kampen (that slow observables can be described by a classical stochastic process) and of Gell-Mann, Hartle, Halliwell and others (that quasi-conserved hydrodynamic modes give rise to decoherent histories). In fact, the spin imbalance operator considered in the main text evolves on a slow time scale (compared to the shortest microscopic evolution time scale determined by the quantum speed limit), which is echoed here as its long time behaviour is dominated by large wavelength ($q/N\approx 0$ or ``quasi-conserved'') hydrodynamic modes. }

\begin{figure}[b]
	\includegraphics[width=1\columnwidth]{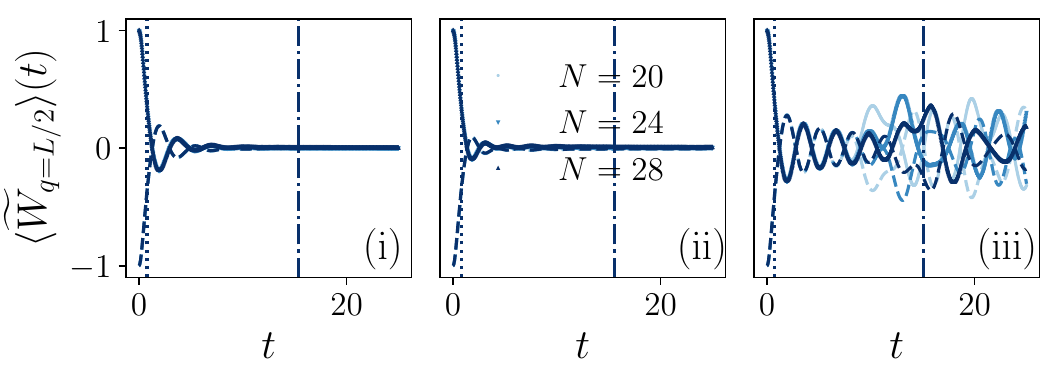}
	\caption{\new{Similar to Fig.~\ref{Pt-DW}, but for $\widetilde{W}_{q=N/2}$. The dotted (dash-dotted) vertical line represents $T= T_\text{neq}$ ($T = T_\text{eq}$), where $T_\text{neq}$ indicates the time at which $\langle \widetilde{W}_{q=N/2}\rangle(t)$ (for largest system size considered $N_\text{max}=28$) decays to $e^{-1}$ of its initial value and $T_\text{eq} = 20 T_\text{neq}$.}}
	\label{Pt-DWF}
\end{figure}

\begin{figure}[t]
	\includegraphics[width=1\columnwidth]{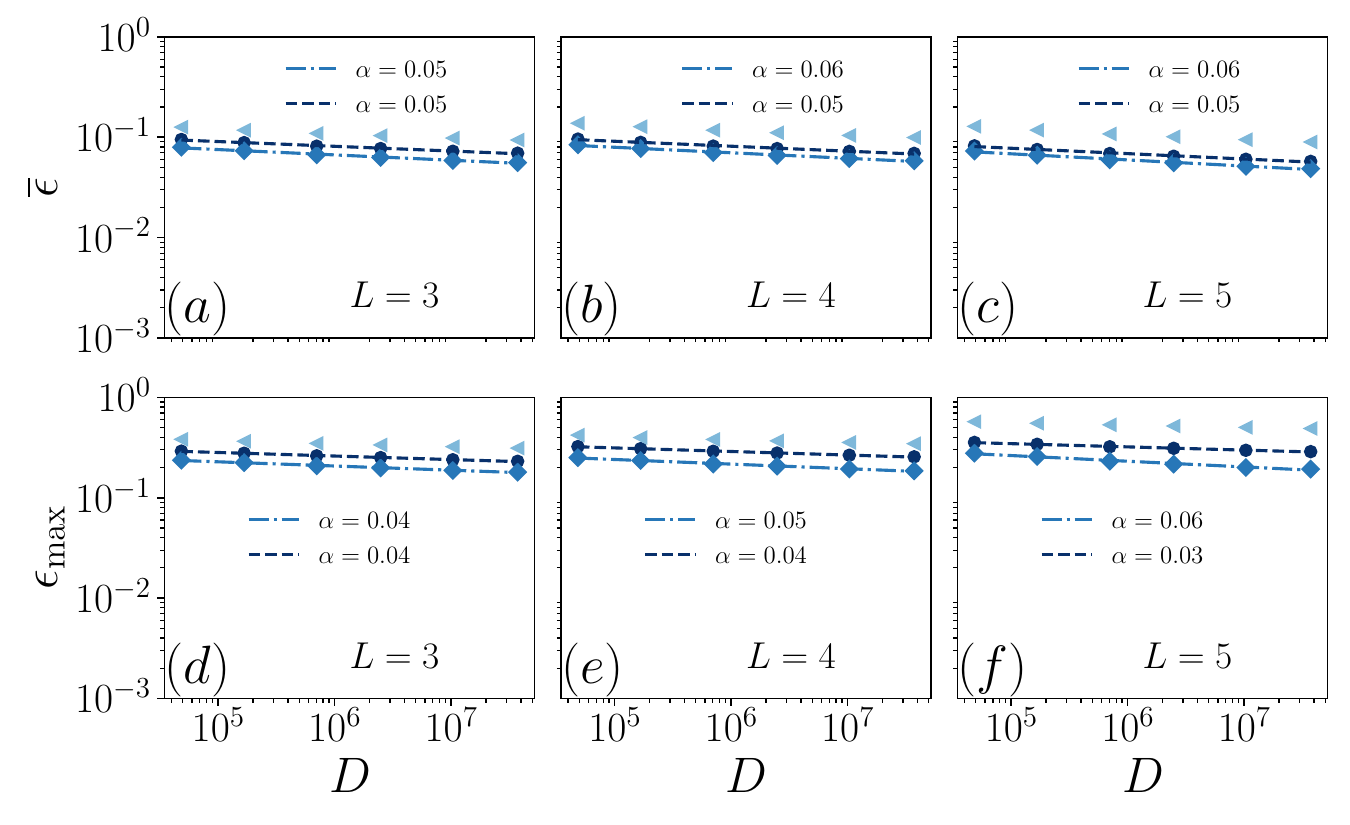}
	\caption{\new{Similar to Fig.~\ref{DF-Short-DWS}, but for $\widetilde{W}_{q=N/2}$.}}
	\label{DF-Short-DWF}
\end{figure}

\new{Furthermore, we also investigate the fast mode \( q = N/2 \). In contrast to the case \( q = 1 \), $ \langle \widetilde{W}_{q=N/2}\rangle (t) $ decays on a much shorter time scale (Fig.~\ref{Pt-DWF}), which is approximately independent of the system size, particularly for the chaotic and interacting integrable case. In other words, $\widetilde{W}_{q=N/2}$ is a fast observable. The emergence of consistent decoherent histories with respect to fast observables is a subtle topic, here we restrict ourselves to reporting the numerical results without delving into detailed discussions.
}

\begin{figure}[b]
	\includegraphics[width=1\columnwidth]{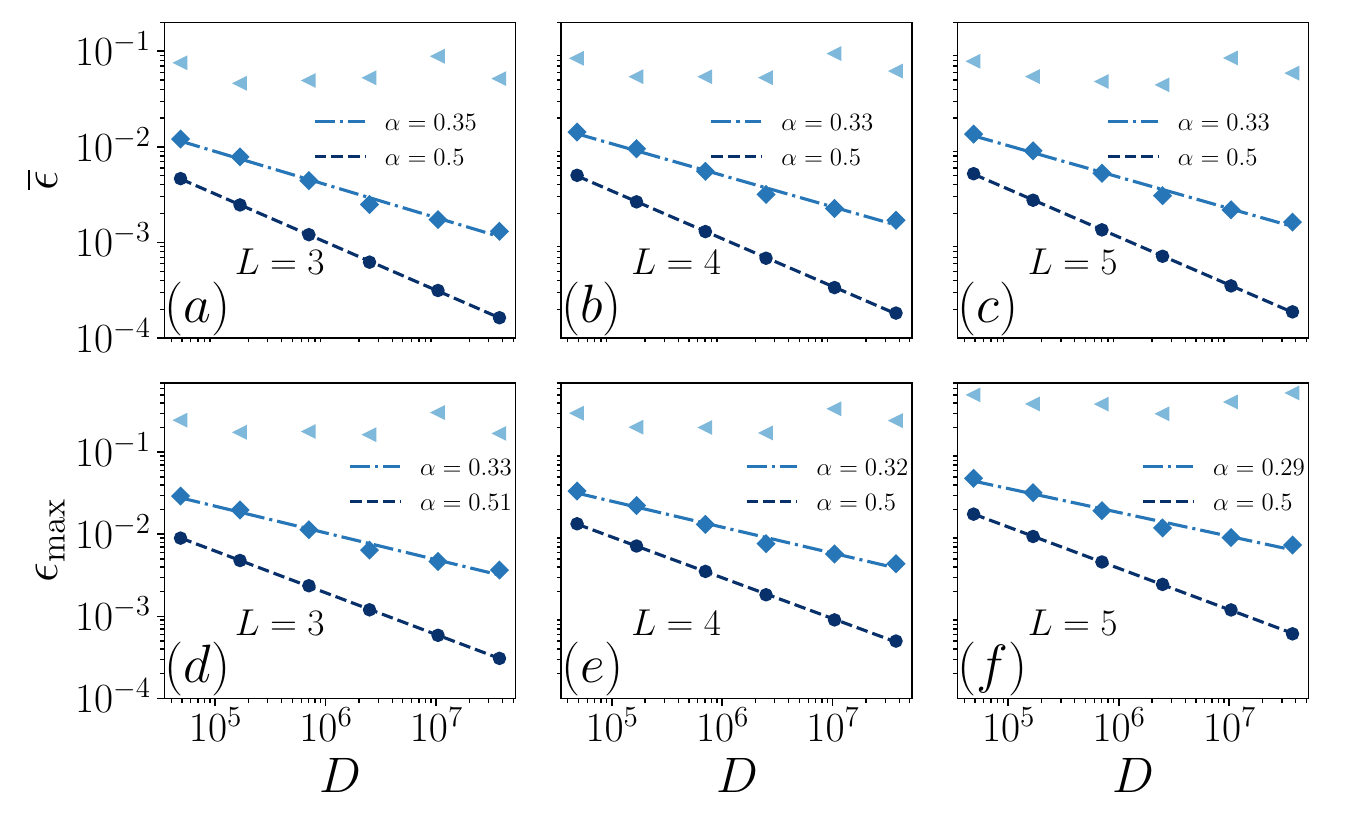}
	\caption{\new{Similar to Fig.~\ref{DF-Long-DWS}, but for $\widetilde{W}_{q=N/2}$.}
	}
	\label{DF-Long-DWF}
\end{figure}

\begin{figure}[t]
	\includegraphics[width=1\columnwidth]{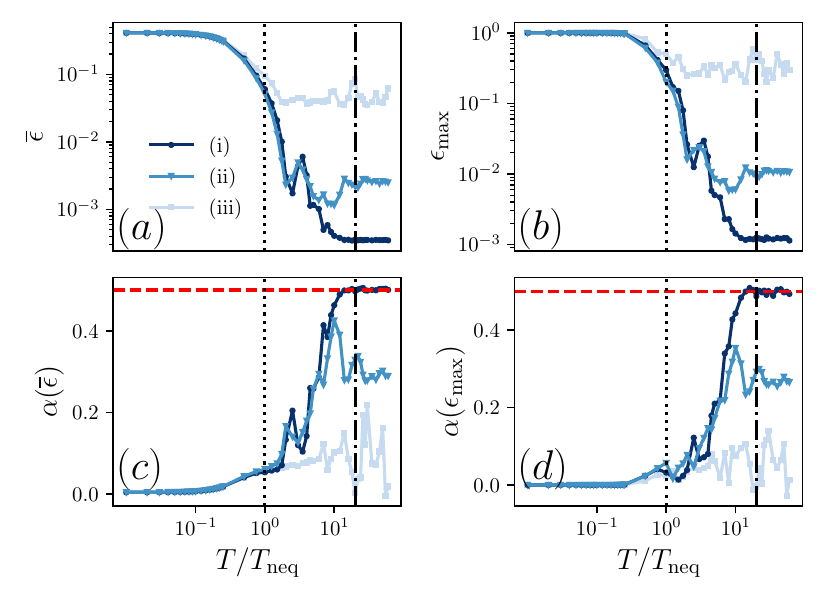}
	\caption{\new{Similar to Fig.~\ref{DF-T-XXZ-DWS}, but for $\widetilde{W}_{q=N/2}$.}}\label{DF-T-XXZ-DWF}
\end{figure}

\new{
	We choose $S_z = 0$ for system size $N = 4k + 2$ and $S_z = 1$ for $N = 4k$, where $k\in \mathbb{N}$. Results are shown in Figs.~\ref{Pt-DWF},~\ref{DF-Short-DWF},~\ref{DF-Long-DWF} and~\ref{DF-T-XXZ-DWF}, which are analogous to Figs.~\ref{Pt-DW},~\ref{DF-Short-DWS},~\ref{DF-Long-DWS} and~\ref{DF-T-XXZ-DWS}. As expected, notably different results are observed compared to the slow mode density wave operator $\widetilde{W}_{q=1}$, especially for the chaotic and interacting integrable case:
	\begin{enumerate}
		\item At the non-equilibrium time scale $T_\text{neq}$ (Fig.~\ref{DF-Short-DWF}), we observe $\alpha \approx 0$ even in the chaotic case indicating a possible sub-exponential or power-law suppression (with respect to $N$) of coherences. Surprisingly, decoherence is even slightly weaker compared to the interacting integrable case.
		\item At the equilibrium time scale $T_\text{eq}$ (Fig.~\ref{DF-Long-DWF}), decoherence is recovered in the chaotic case ($\alpha \approx 0.5$), and a similar behavior appears to hold in the interacting integrable case with a smaller $\alpha \approx 0.3$.
		\item The overall behavior of the scaling exponent $\alpha$ as a function of $T$ (Fig.~\ref{DF-T-XXZ-DWF}) is similar to the slow mode except for two differences. First, the onset of decoherence happens later for all cases. Second, the scaling exponent in case (ii) remains finite ($\alpha \approx 0.25$) and does not decay to zero.
	\end{enumerate}
	To conclude, we found a noticeably different behaviour of the fast mode compared to the three slow cases (the spin imbalance in the main text, the slow $q=1$ mode above and the example in the next section). In particular, our results show that coarseness alone is not sufficient to guarantee decoherence for nonequilibrium time scales, even in a chaotic system. However, in unison with the three slow cases we also found here a distinct behavior for the (i) chaotic, (ii) interacting integrable, and (iii) non-interacting integrable cases for large times. }

\subsection*{S2. Numerical results in Ising model}

To analyze the generality of our main results, we also consider an Ising chain with Hamiltonian
\begin{equation}
H=\sum_{\ell=1}^{N}\left(h_{x}\sigma_{x}^{\ell}+h_{z}\sigma_{z}^{\ell}+J\sigma_{z}^{\ell}\sigma_{z}^{\ell}\right).
\end{equation}
We assume periodic boundary condition and set $J=h_x=1.0$. Two different values of $h_z$ are considered: (i) for $h_z = 0.5$ (titled field) the system is chaotic; (iii) for $h_z = 0.0$ (transverse field) the system is integrable and can be mapped to free fermions. \new{Note that the system is not known to have an interacting integrable case (ii).}
A natural operator of interest here is an energy imbalance operator,
\begin{equation}
B_0 = H^L - H^R,
\end{equation}
where
\begin{gather}
H^{L}=\sum_{\ell=1}^{\frac{N}{2}}(h_{x}\sigma_{x}^{\ell}+h_{z}\sigma_{z}^{\ell})+\sum_{\ell=1}^{\frac{N}{2}-1}J\sigma_{z}^{\ell}\sigma_{z}^{\ell+1}, \nonumber \\
H^{R}=\sum_{\ell=\frac{N}{2}+1}^{N}(h_{x}\sigma_{x}^{\ell}+h_{z}\sigma_{z}^{\ell})+\sum_{\ell=\frac{N}{2}+1}^{N-1}J\sigma_{z}^{\ell}\sigma_{z}^{\ell+1}.
\end{gather}
It quantifies an ``energy bias'' between the left and right half of the spin chain.
Denoting its eigenvectors and eigenvalues by $B_0|b_k\rangle = b_k|b_k\rangle$, we construct a coarse observable $B = \Pi^B_{+}-\Pi^B_{-}$ with projectors
\be
\Pi^B_{+}=\sum_{b_{k}>0}|b_{k}\rangle\langle b_{k}|\quad\text{and}\quad\Pi^B_{-}=\sum_{b_{k}\le0}|b_{k}\rangle\langle b_{k}|.
\ee
The subspaces are denoted by ${\cal H}^B_\pm$, with Hilbert space dimension $V^B_\pm = \text{dim} {\cal H^B_\pm} =  \text{Tr}[\Pi^B_\pm]$. In this model, $V^B_+ \approx V^B_-$.

\begin{figure}[t]
	\includegraphics[width=1\columnwidth]{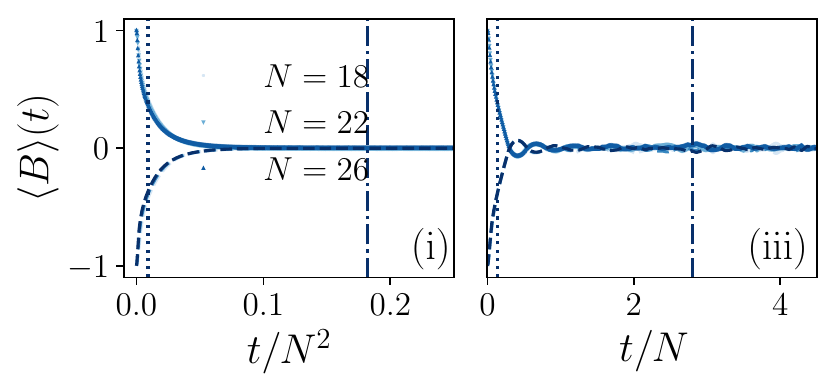}
	\caption{The expectation value $\langle B\rangle(t)$ of the coarse energy imbalance in the Ising model as a function of rescaled time for initial states $|\psi^+_0\rangle$ (solid line) and $|\psi^-_0\rangle$ (dashed line) for chaotic (i) and free (iii) cases. The dotted (dash-dotted) vertical line represents $T= T_\text{neq}$ ($T = T_\text{eq}$), where $T_\text{neq}$ indicates the time at which $\langle B\rangle(t)$ decays to $e^{-1}$ of its initial value and $T_\text{eq} = 20 T_\text{neq}$. Specifically, we obtain $T_\text{neq}$ from the data of the largest system size $(N_\text{max} = 26)$. Then, for smaller sizes $N$ we set $T_{\text{neq}}(N)=(N/N_{\text{max}})^{p}T_{\text{neq}}(N_{\text{max}})$ with $p=2$ for (i) and $p = 1$ for (iii). \new{The free case is denoted as (iii) to remain consistent with the notation used in the main text.}
	}
	\label{Pt-Ising}
\end{figure}

As a start, we plot the expectation value $\langle B\rangle(t)$ of the coarse energy imbalance as a function of time in Fig.~\ref{Pt-Ising}. This is done for two different non-equilibrium initial states $|\psi_0^{\pm}\rangle$, where $|\psi_0^\pm\rangle$ is a Haar random state restricted to the subspace $\C H^B_\pm$. In the chaotic case (i) the system relaxes to its thermal equilibrium value $\langle B\rangle_\text{eq} = 0$ with an equilibration time scale $\propto N^2$. In contrast, in the free case (iii) $\langle B\rangle(t)$ decays on a time scale $\propto N$ and fluctuations around $\langle B\rangle_\text{eq}$ remain visible for all times that we considered. Note, however, that---compared to the free case of the Heisenberg chain considered in the main text [cf.~Fig.~1]---equilibration seems to work much better for the free case of the Ising model.

\begin{figure}[t]
	\includegraphics[width=1\columnwidth]{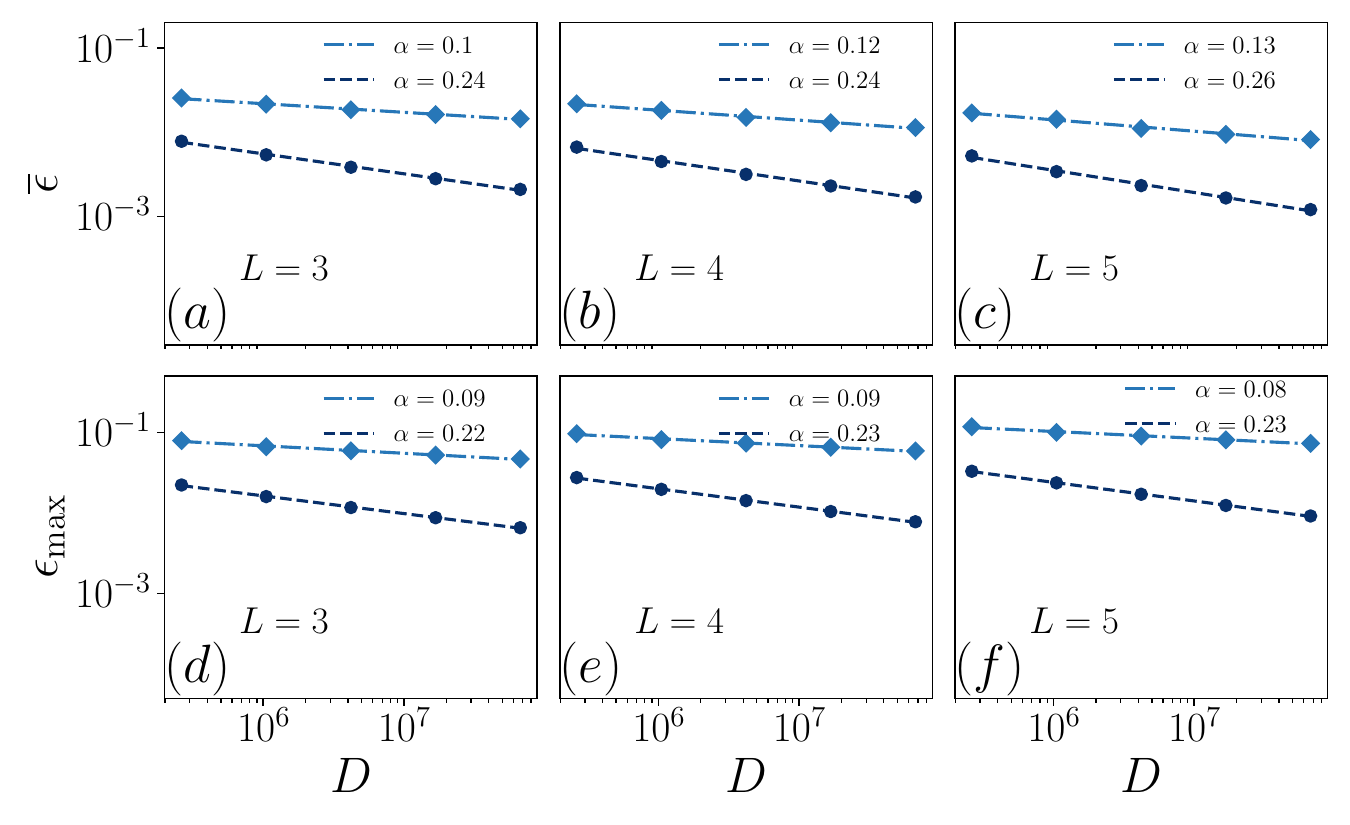}
	
	\caption{Average $\overline{\epsilon}$ and maximum $\epsilon_\text{max}$ amount of coherence versus Hilbert space dimension $D$ of the coarse energy imbalance in Ising model for the (i) chaotic (dark blue disks) and (ii) interacting integrable (medium blue diamonds) case. The dashed and dash-dotted line fit a scaling law of the form $D^{-\alpha}$ to (i) and (ii). The time step is $T = T_\text{neq}$ and the system sizes are $N=18,20,\dots, 26$.
	}
	\label{DF-Ising-Short}
\end{figure}

\begin{figure}[b]
	\includegraphics[width=1\columnwidth]{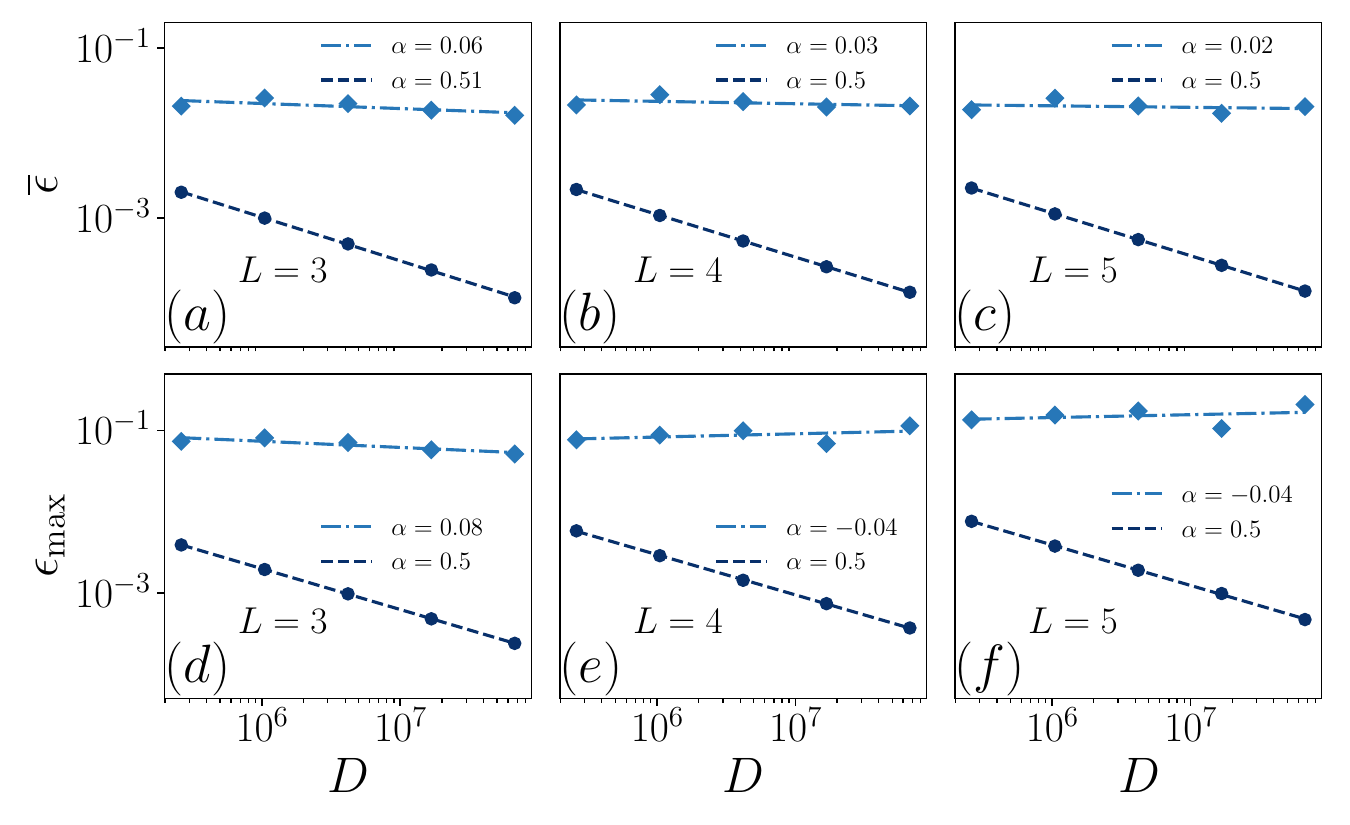}
	
	\caption{Identical to Fig. \ref{DF-Ising-Short} except for time steps $T = T_\text{eq}$.
	}
	\label{DF-Ising-Long}
\end{figure}

The emergence of decoherence is investigated in Figs.~\ref{DF-Ising-Short} and \ref{DF-Ising-Long} for Haar random initial states $|\psi_0\rangle$. We plot in double logarithmic scale $\overline{\epsilon}$ and $\epsilon_\text{max}$ versus the Hilbert space dimension $D$ for histories of lengths $L\in\{3,4,5\}$  for two different $T$: a nonequilibrium time scale $T_\text{neq}$ in Fig.~\ref{DF-Ising-Short} (identical to the dotted line in Fig.~\ref{Pt-Ising}) and an equilibrium time scale $T_\text{eq}$ in Fig.~\ref{DF-Ising-Long} (identical to the dash-dotted line in Fig.~\ref{Pt-Ising}).  Finally, each data point in Figs.~\ref{DF-Ising-Short} and \ref{DF-Ising-Long} is obtained by averaging $\overline{\epsilon}$ and $\epsilon_\text{max}$ over $2^{30-N}$ different realizations of $|\psi_0\rangle$.

%

\begin{figure}[t]
	\includegraphics[width=1\columnwidth]{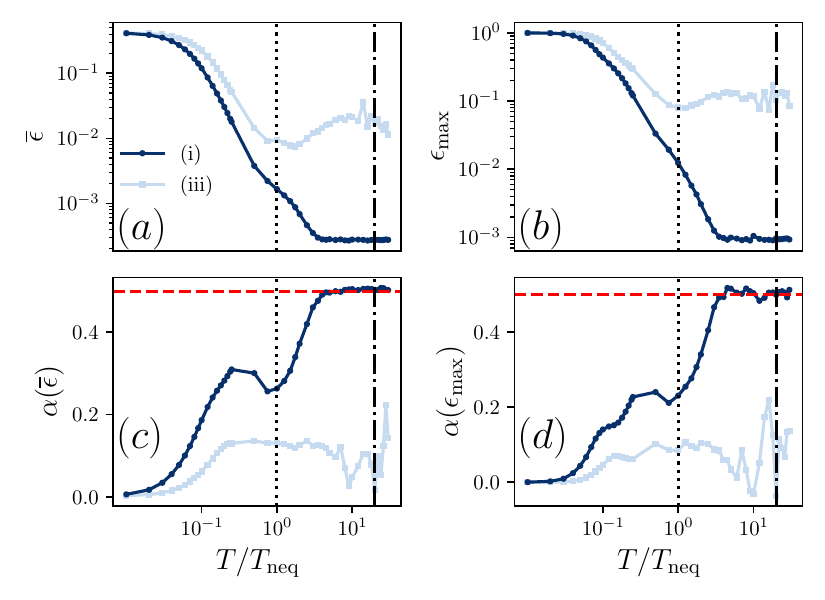}
	\caption{Decoherence as a function of the time step $T = t_n-t_{n-1}$ for history length $L=5$  of the energy imbalance in the Ising model for the (i) chaotic (dark blue) and the (iii) free (light blue) cases. (a,b) $\overline{\epsilon}, \epsilon_\text{max}$ for $N = 24$ on a double logarithmic scale. (c,d) Scaling exponent $\alpha$ for $\overline{\epsilon}, \epsilon_\text{max}$ obtained for system sizes $N = 18,20,\ldots,26$ and as a mean over $2^{28-N}$ realizations of a Haar random $|\psi_0\rangle$. \new{Black dotted (dash-dotted) line indicates $T=T_\text{neq} (T=T_\text{eq})$.} Red dashed line in (c,d) indicates $\alpha = 0.5$.}
	\label{Epsilon-T-Ising}
\end{figure}

For the chaotic case (i) we find a scaling law of the form $D^{-\alpha}$ (note that $D\propto 2^{N}$) with \new{$\alpha \approx 0.25$} at the nonequilibrium time scale and with $\alpha\approx 0.5$ at the equilibrium time scale (with the same $\alpha$ for both $\overline{\epsilon}$ and $\epsilon_\text{max}$). This again indicates a robust exponential suppression (with respective to system size $N$) of coherences in chaotic systems. For the free case (iii) we find smaller exponents $\alpha\approx 0.1$ at the nonequilibrium time scale, and $\alpha \approx 0$ at the equilibrium time scale.


\new{Furthermore, similar to Fig.~4 in the main text, we plot $\overline{\epsilon}$ and $\epsilon_\text{max}$ for system size $N = 24$ as well as the scaling exponent $\alpha(\overline{\epsilon})$ ($\alpha({\epsilon}_{\text{max}})$) versus $T$ for history length $L = 5$ in the Ising model in Fig.~\ref{Epsilon-T-Ising}. Similar trends are observed: in chaotic case, both $\overline{\epsilon}$ and $\epsilon_\text{max}$ decrease with increasing $T$. In contrast, in the free case, they increase on average (after a sudden drop at small $T$), accompanied by significant fluctuations.
	The general behavior of the scaling exponent $\alpha(\overline{\epsilon})$ ($\alpha({\epsilon}_{\text{nax}})$), especially for time scale $T\ge T_{\text{neq}}$, is also similar to the results shown in Fig.~4. This cements the idea that the decoherence behavior of slow observables is in general characterized by common qualitative features that clearly distinguish between integrable and non-integrable dynamics.
}

\subsection*{S3. Level statistics}

To study the chaoticity (integrability) of the considered models, we analyze the distribution of the nearest-level spacing of the unfolded spectrum.
After unfolding, the averaged level density becomes constant (usually set to $1$, as is done here).  The ordered eigenvalues of the unfolded spectrum are denoted by $\tilde{E}_i$.
As an indicator of quantum chaos, we study the distribution of $s_i = \tilde{E}_{i+1} - \tilde{E}_{i}$, denoted by $P(s)$.
$P(s)$ differentiates between chaotic and integrable systems: i) For chaotic systems, $P(s)$ follows a Wigner-Dyson distribution, where for systems with time-reversal symmetry,
\begin{equation}\label{eq-PS-GOE}
P(s)=P_\text{GOE}(s)=\frac{\pi}{2}se^{-\frac{\pi}{4}s^{2}},
\end{equation}
which is the prediction of Gaussian Orthogonal Ensemble (GOE);
ii) For integrable systems, $P(s)$ follows a Poisson distribution
\begin{equation}\label{eq-PS-Poisson}
P(s) = P_\text{Poisson}(s) = e^{-s}\ .
\end{equation}

\begin{figure}[t]
	\includegraphics[width=1.0\columnwidth]{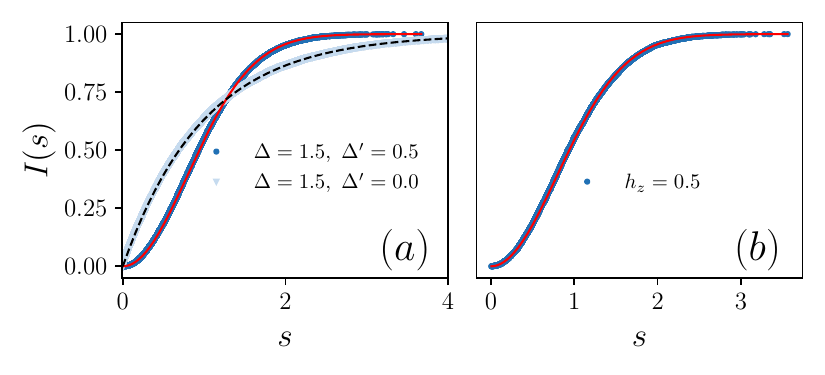}
	
	\caption{Level statistics: cumulative distribution  of the nearest-level spacing $I(s)$ versus $s$, for (a) XXZ model ($N = 24$) and (b) Ising model ($N = 20$).
		The solid and dashed line indicates $I_\text{GOE}(s)$ and $I_\text{Poisson} (s)$, respectively (Eq.~\eqref{eq-Is}).
	}
	\label{Fig-PS}
\end{figure}

In practise, we consider the cumulative distribution of $P(s)$
\begin{equation}
I(s) = \int _0 ^s P(r) dr,
\end{equation}
and compare it to
\begin{equation}\label{eq-Is}
I_\text{GOE}(s) = 1 - \exp{(-\frac{\pi}{4}s^2)},\ I_\text{Poisson}(s) = 1 - \exp{(-s)}.
\end{equation}
In both models, due to the existence of additional global symmetries (translational invariance, reflection invariance, etc.) alongside with total energy conservation, our analysis is confined to a specific subspace. We compute the $P(s)$ by considering $1/2$ of the total eigenvalues located in the middle of the spectrum, and results are shown in Fig.~\ref{Fig-PS}. Good agreement with the Wigner-Dyson distribution  is observed in the XXZ model ($\Delta_1 = 1.5,\ \Delta_2 = 0.5$) and in the Ising model ($h_z = 0.5$), indicating the systems are chaotic with respect to the corresponding parameter. In contrast, in  XXZ model ($\Delta_1 = 1.5,\ \Delta_2 = 0.0$), a Possion distribution is found, which suggests that the system is integrable. The results are in line with our findings in the main text. The results for trivial cases, where the model is equivalent to free fermions, e.g., XXZ model ($\Delta_1 = \Delta_2 = 0$) and Ising model $(h_z = 0.0)$, are not shown here.

\end{document}